\documentclass[times, 10pt,twocolumn]{article} 
\usepackage{latex8}

\usepackage{url}
\usepackage{cite}
\usepackage{color}
\usepackage{epsfig}
\usepackage{amsmath, amssymb, amsfonts}
\usepackage{graphicx}
\usepackage{subfigure}
\usepackage{balance}
\usepackage{verbatim}
\usepackage{algorithm}
\usepackage{algorithmic}
\usepackage{times}
\usepackage{float}
\usepackage{multirow}
\usepackage{eurosym}

\newcommand{\tabincell}[2]{\begin{tabular}{@{}#1@{}}#2\end{tabular}}
\newcommand{\para}[1]{{\vspace{4pt} \bf \noindent #1 \hspace{10pt}}}

\newcommand{\compact }{\vspace{-0.1in}}

\newenvironment{packed_itemize}{
\begin{list}{\labelitemi}{\leftmargin=1.5em}
  \setlength{\itemsep}{1pt}
  \setlength{\parskip}{0pt}
  \setlength{\parsep}{0pt}
  \setlength{\headsep}{0pt}
  \setlength{\topskip}{0pt}
  \setlength{\topmargin}{0pt}
  \setlength{\topsep}{0pt}
  \setlength{\partopsep}{0pt}  
}{\end{list}}

\newcommand{\eg}{{\it e.g.\ }}
\newcommand{\etal}{{\it et al.\ }}
\newcommand{\etc}{{\it etc}}
\newcommand{\ie}{{\it i.e.\ }}

\begin{document}

\title{Social Turing Tests: Crowdsourcing Sybil Detection}

\author{Gang Wang, Manish Mohanlal, Christo Wilson, Xiao Wang$^\ddag$, \\
	Miriam Metzger$^{\dag}$, Haitao Zheng and Ben Y. Zhao \\
        \\
        Department of Computer Science, U. C. Santa Barbara, CA USA\\
	$^\dag$Department of Communications, U. C. Santa Barbara, CA USA\\
	$^\ddag$Renren Inc., Beijing, China \\
}

\maketitle
\thispagestyle{empty}

\begin{abstract}
  As popular tools for spreading spam and malware, Sybils (or fake accounts)
  pose a serious threat to online communities such as Online Social Networks
  (OSNs). Today, sophisticated attackers are creating realistic Sybils that
  effectively befriend legitimate users, rendering most automated Sybil
  detection techniques ineffective.  In this paper, we explore the
  feasibility of a {\em crowdsourced} Sybil detection system for OSNs. We
  conduct a large user study on the ability of humans to detect today's Sybil
  accounts, using a large corpus of ground-truth Sybil accounts from the
  Facebook and Renren networks.  We analyze detection accuracy by both
  ``experts'' and ``turkers'' under a variety of conditions, and find that
  while turkers vary significantly in their effectiveness, experts
  consistently produce near-optimal results.  We use these results to drive
  the design of a multi-tier crowdsourcing Sybil detection system.  Using our
  user study data, we show that this system is scalable, and can be highly
  effective either as a standalone system or as a complementary technique to
  current tools.
 
\end{abstract}

\section{Introduction}
\label{sec:intro}
\compact

The rapid growth of Sybil accounts is threatening the stability and security
of online communities, particularly online social networks (OSNs).  Sybil
accounts represent fake identities that are often controlled by a small
number of real users, and are increasingly used in coordinated campaigns to
spread spam and malware~\cite{crowdturf-www12,osnspam-imc10}.  In fact,
measurement studies have detected hundreds of thousands of Sybil
accounts in different OSNs around the world~\cite{sybilrank,sybils-imc11}.
Recently, Facebook revealed that up to 83 million of its users may be
fake\footnote{\url{http://www.bbc.com/news/technology-19093078}}, up
significantly from 54 million earlier\footnote{\url{http://www.bbc.com/news/technology-18813237}}.

The research community has produced a substantial number of techniques for
automated detection of Sybils~\cite{sybilguard,sybillimit,sybilinfer}.
However, with the exception of SybilRank~\cite{sybilrank}, few have been
successfully deployed. The majority of these techniques rely on the assumption that
Sybil accounts have difficulty friending legitimate users, and thus tend to
form their own communities, making them visible to community detection
techniques applied to the social graph~\cite{sybilcommunity}. 

Unfortunately, the success of these detection schemes is likely to decrease
over time as Sybils adopt more sophisticated strategies to ensnare legitimate
users. 
First, early user studies on OSNs such as Facebook show that users are often
careless about who they accept friendship requests from~\cite{socialbot-acsac}. Second,
despite the discovery of Sybil communities in Tuenti~\cite{sybilrank}, not
all Sybils band together to form connected components. For example, a recent
study of half a million Sybils on the Renren network~\cite{renren-imc10} showed
that Sybils rarely created links to other Sybils, and instead intentionally try to
infiltrate communities of legitimate users~\cite{sybils-imc11}.  Thus,
these Sybils rarely connect to each other, and do not form communities. Finally, there
is evidence that creators of Sybil accounts are using advanced
techniques to create more realistic profiles, either by copying
profile data from existing accounts, or by recruiting real users to customize
them~\cite{crowdturf-www12}. Malicious parties are willing to pay for these
authentic-looking accounts to better befriend real users.

These observations motivate us to search for a new approach to detecting
Sybil accounts.  Our insight is that while attackers are creating more
``human'' Sybil accounts, fooling intelligent users, {\em i.e.}  passing a
``social Turing test,'' is still a very difficult task.  Careful users can
apply intuition to detect even small inconsistencies or discrepancies in the
details of a user profile.  Most online communities already have mechanisms
for users to ``flag'' questionable users or content, and social networks
often employ specialists dedicated to identifying malicious content and
users~\cite{sybilrank}.  While these mechanisms are ad hoc and costly, our
goal is to explore a scalable and systematic approach of applying human
effort, {\em i.e.}  crowdsourcing, as a tool to detect Sybil accounts.

Designing a successful crowdsourced Sybil detection requires that we first
answer fundamental questions on issues of accuracy, cost, and scale. One key
question is, how accurate are users at detecting fake accounts? More
specifically, how is accuracy impacted by factors such as the user's
experience with social networks, user motivation, fatigue, and
language and cultural barriers?  Second, how much would it cost to
crowdsource authenticity checks for all suspicious profiles? Finally, how can
we design a crowdsourced Sybil detection system that scales to millions of
profiles?

In this paper, we describe the results of a large user study into the
feasibility of crowdsourced Sybil detection.  We gather ground-truth data
on Sybil accounts from {\em three} social network populations:
Renren~\cite{renren-imc10}, the largest social network in China, Facebook-US,
with profiles of English speaking users, and Facebook-India, with profiles of
users who reside in India.  The security team at Renren Inc. provided us with
Renren Sybil account data, and we obtained Facebook (US and India) Sybil
accounts by crawling highly suspicious profiles weeks before they were banned by
Facebook. Using this data, we perform user studies analyzing the effectiveness
of Sybil detection by three user populations: motivated and experienced
``experts''; crowdsourced workers from China, US, and India; and a group
of UCSB undergraduates from the Department of Communications.

Our study makes three key contributions. First, we analyze detection
accuracy across different datasets, as well as the impact of different
factors such as demographics, survey fatigue, and OSN experience.
We found that well-motivated experts and undergraduate
students produced exceptionally good detection rates with near-zero false positives.
Not surprisingly, crowdsourced workers missed more Sybil accounts, but still
produced near zero false positives. We observe that as testers examine more
and more suspicious profiles, the time spent examining each profile decreases.
However, experts maintained their accuracy over time while crowdworkers made more
mistakes with additional profiles. Second, we performed detailed analysis on individual
testers and account profiles. We found that while it was easy to identify a
subset of consistently accurate testers, there were very few ``chameleon profiles''
that were undetectable by all test groups.  Finally, we propose a scalable
crowdsourced Sybil detection system based on our results, and use
trace-driven data to show that it achieves both accuracy and scalability with
reasonable costs.

By all measures, Sybil identities and fake accounts are growing rapidly on
today's OSNs. Attackers continue to innovate and find better ways of
mass-producing fake profiles, and detection systems must keep
up both in terms of accuracy and scale. This work is the first to propose
crowdsourcing Sybil detection, and our user study results are extremely
positive. We hope this will pave the way towards testing and deployment
of crowdsourced Sybil detection systems by large social networks.

\section{Background and Motivation}
\label{sec:background}

Our goal is to motivate and design a crowdsourced Sybil
detection system for OSNs. First, we briefly introduce the concept of
crowdsourcing and define key terms. Next, we review the current state of social
Sybil detection, and highlight ongoing challenges in this area. Finally, we
introduce our proposal for crowdsourced Sybil detection, and enumerate the key
challenges to our approach.

\subsection{Crowdsourcing}
\compact

Crowdsourcing is a process where work is outsourced to an undefined group of
people. The web greatly simplifies the task of gathering virtual groups of
workers, as demonstrated by successful projects such as Wikipedia.
Crowdsourcing works for any job that can be decomposed into short, simple
tasks, and brings significant benefits to tasks not easily performed by
automated algorithms or systems.
First, by harnessing small amounts of work from many people, no individual is
overburdened. Second, the group of workers can change dynamically, which
alleviates the need for a dedicated workforce. Third, workers can be
recruited quickly and on-demand, enabling elasticity.
Finally and most importantly, by leveraging human intelligence, crowdsourcing
can solve problems that automated techniques cannot.

In recent years, crowdsourcing websites have emerged that allow anyone to
post small jobs
to the web and have them be solved by crowdworkers for a small fee. The pioneer in
the area is Amazon's Mechanical Turk, or {\em MTurk} for short. On MTurk, anyone can
post jobs called Human Intelligence tasks, or {\em HITs}.  Crowdworkers
on MTurk, or {\em turkers}, complete HITs and collect the
associated fees. Today, there are around 100,000 HITs available on MTurk
at any time, with 90\% paying $\le$\$0.10 each~\cite{amazonanalysis-acm, demographic-chi10}.
There are over 400,000 registered turkers on MTurk, with 56\% from the US,
and 36\% from India~\cite{demographic-chi10}.

Social networks have started to leverage crowdsourcing to augment their workforce. For
example, Facebook crowdsources content moderation tasks, including filtering pornographic and violent pictures and videos~\cite{oDeskface}. However, to
date we know of no OSN that
crowdsources the identification of fake accounts. Instead, OSNs like Facebook and Tuenti
maintain dedicated, in-house staff for this purpose~\cite{oDeskface, sybilrank}.

Unfortunately, attackers have also begun to leverage crowdsourcing.  Two
recent studies have uncovered crowdsourcing websites where malicious users
pay crowdworkers to create Sybil accounts on OSNs and generate
spam~\cite{crowdturf-www12, freelance}. These Sybils are particularly
dangerous because they are created and managed by real human beings, and thus
appear more authentic than those created by automated scripts.  Crowdsourced
Sybils can also bypass traditional security mechanisms, such as CAPTCHAs,
that are designed to defend against automated attacks.

\subsection{Sybil Detection}
\compact

The research community has produced many systems designed to detect Sybils on
OSNs. However, each one relies on specific assumptions about Sybil behavior and
graph structure in order to function. Thus, none of these systems is general
enough to perform well on all OSNs, or against Sybils using different attack
strategies.

The majority of social Sybil detectors from the literature rely on two key
assumptions. First, they assume that Sybils have trouble friending legitimate
users.  Second, they assume that Sybil accounts create many edges amongst
themselves.  This leads to the formation of well-defined Sybil communities that have
a small quotient-cut from the honest region of the graph~\cite{sybilguard, sybillimit,
sumup,sybilcommunity,sybilinfer}.  Although similar Sybil
community detectors have been shown to work well on the Tuenti OSN~\cite{sybilrank},
other studies have demonstrated limitations of this approach. For example,
a study by Yang et al. showed that Sybils on the Renren OSN do not form connected
components at all~\cite{sybils-imc11}.  Similarly, a meta-study of multiple OSN graphs
revealed that many are not fast-mixing, which is a necessary precondition for
Sybil community detectors to perform well~\cite{mixing-imc10}.

Other researchers have focused on feature-based Sybil detectors. Yang et al.
detect Sybils by looking for accounts that send many friend requests that are
rejected by the recipient. This detection technique works well on Renren because
Sybils must first attempt to friend many users before they can begin effectively
spamming~\cite{sybils-imc11}. However, this technique does not generalize. For example,
Sybils on Twitter do not need to create social connections, and instead send
spam directly to any user using ``@'' messages.

Current Sybil detectors rely on Sybil behavior assumptions that make them
vulnerable to sophisticated attack strategies. For example, Irani et al.
demonstrate that ``honeypot'' Sybils are capable of passively gathering
legitimate friends and penetrating the social
graph~\cite{honeysybil-dimva11}.  Similarly, some attackers pay users to
create fake profiles that bypass current detection
methods~\cite{crowdturf-www12}.  As Sybil creators adopt more sophisticated
strategies, current techniques are likely to become less effective.

\subsection{Crowdsourcing Sybil Detection}
\compact

In this study, we propose a crowdsourced Sybil detection system. We believe this
approach is promising for three reasons: first, humans can make overall judgments
about OSN profiles that are too complex for automated algorithms. For example,
humans can evaluate the sincerity of photographs and understand subtle conversational
nuances. Second, social-Turing tests are resilient to changing attacker strategies,
because they are not reliant on specific features. Third, crowdsourcing is much
cheaper than hiring full-time content moderators~\cite{snow-emnlp08,Heer-chi10}.
However, there are several questions that we must answer to verify that this
system will work in practice:

\begin{packed_itemize}
   \item How {\bf accurate} are users at distinguishing between real and fake profiles?
   Trained content moderators can perform this task, but can crowdworkers
   achieve comparable results?
   \item Are there {\bf demographic factors} that affect detection accuracy? Factors like
   age, education level, and OSN experience may impact the accuracy of crowdworkers.
   \item Does {\bf survey fatigue} impact detection accuracy? In many instances, people's
   accuracy at a task decline over time as they become tired and bored.
   \item Is crowdsourced Sybil detection {\bf cost effective}? Can the system be scaled to
   handle OSNs with hundreds of millions of users?
\end{packed_itemize}

We answer these questions in the following sections. Then, in
Section~\ref{sec:prac}, we describe the design of
our crowdsourced Sybil detection system, and use our user data to validate
its effectiveness.

\section{Experimental Methodology}
\label{sec:exp}
\compact

In this section, we present the design of our user studies to validate
the feasibility of crowdsourced Sybil detection. First, we introduce the
three datasets used in our experiments: Renren, Facebook US, and Facebook
India. We describe how each dataset was gathered, and how the ground-truth
classification of Sybil and legitimate profiles was achieved. Next, we describe
the high-level design of our user study and its website implementation.
Finally, we introduce the seven groups of test subjects. Test subjects are
grouped into experts, turkers from crowdsourcing websites, and university
undergraduates. We use different test groups from China, the US, and India
that correspond to our three datasets. All of our data collection and
experimental methodology was evaluated and received IRB approval before
we commenced our study.

\subsection{Ground-truth Data Collection}
\compact

Our experimental datasets are collected from two large OSNs: Facebook 
and Renren. Facebook is the most popular OSN in the world and has more 
than 1 billion users~\cite{facebook900}. Renren is the largest OSN in China, with more than 220 million 
users~\cite{renren-imc10}. Both sites use similar visual layouts and 
offer user profiles with similar features, including space for basic 
information, message ``walls,'' and photo albums.  Basic information 
in a profile includes items like name, gender, a profile image, 
total number of friends, interests, \etc.

Each dataset is composed of three types of user profiles: confirmed {\em Sybils},
confirmed {\em legitimate} users, and {\em suspicious} profiles that are likely to be
Sybils. Confirmed Sybil profiles are known to be fake because they have been
banned by the OSN in question, and manually verified by us. Suspicious profiles exhibit
characteristics
that are highly indicative of a Sybil, but have not been banned by the OSN. Legitimate
profiles have been hand selected and verified by us to ensure their integrity.
We now describe the details of our data collection process on Facebook
and Renren.

\begin{figure}[t]
	\includegraphics[width=0.45\textwidth]{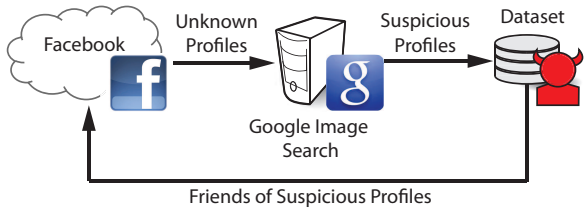}
	\vspace{-0.1in}
	\caption{Facebook crawling methodology. }
	\label{fig:crawler}
    \vspace{-0.1in}
\end{figure}

\para{Facebook.} We collect data from Facebook using a custom web crawler. Because
Facebook caters to an international audience, we specifically targeted two regional
areas for study: the US and India. We chose these two regions because they have large,
Internet enabled populations, and both countries have active marketplaces for
crowdworkers~\cite{demographic-chi10}. Our Facebook crawls were conducted between December 2011 and January 2012.

The legitimate profiles for our study were randomly selected from a pool of 86K profiles.
To gather this pool of profiles, we seeded our crawler with 8 Facebook profiles
belonging to members of our lab (4 in the US, and 4 in India). The crawler then
visited each seed's friends-of-friends, \ie the users two-hops away on the
social graph.
Studies have shown that trust on social networks is often transitive~\cite{trust}, and
thus the friends-of-friends of our trusted seeds are likely to be trustworthy as well.
From the 86K total friends-of-friends in this set, the crawler sampled 100 profiles
(50 from the US, 50 from India) that had Facebook's default, permissive privacy settings.
We manually examined all 100 profiles to make sure they were 1) actually legitimate users,
and 2) we did not know any of them personally (to prevent bias in our
study). 

To facilitate collection of Sybils on Facebook, we make one
assumptions about Sybil behavior: we assume that Sybils use {\em widely available
photographs} from the web as profile images. Intuitively, Sybils need realistic profile images
in order to appear legitimate. Hence, Sybils must resort to using publicly available images
from around the web. Although {\em all} Sybils on Facebook may not obey this assumption,
we will show that enough do to form a sufficiently large sample for our user study.

To gather suspicious profiles, we seeded our crawler with the profiles of known Sybils
on Facebook~\cite{fakeaccountsartice}. The crawler then snowball crawled outward from
the initial seeds. We leveraged {\em Google Search by Image} to locate profiles using
widely available photographs as profile images. Figure~\ref{fig:crawler} illustrates this process.
For each profile visited by the crawler, {\em all} of its profile images were sent to
Google Search by Image (Facebook maintains a photo album for each user that includes
their current profile image, as well as all prior images). If Google Search by Image
indexed $\ge$90\% of the profile images on sites other than Facebook, then we consider
the account to be suspicious. The crawler recorded the basic information, wall, and photo
albums from each suspicious profile. We terminated the crawl after a sufficient number
of suspicious profiles had been located.

We search for all of a user's profile images rather than just the current image
because legitimate users sometimes use stock photographs on their profile
(\eg a picture of their favorite movie star). We eliminate these false positives
by setting minimum thresholds for suspicion: we only consider profiles with $\ge$2 profile images,
and if $\ge$90\% are available on the web, then the profile is considered suspicious.

In total, our crawler was able to locate 8779 suspicious Facebook profiles. Informal,
manual inspection of the profile images used by these accounts reveals that most use
pictures of ordinary (usually attractive) people. Only a small number of accounts use
images of recognizable celebrities or non-people (\eg sports cars). Thus, the majority
of profile images in our dataset are not suspicious at first-glance. Only by using
external information from Google does it become apparent that these photographs have
been misappropriated from around the web.

At this point, we don't have ground-truth about these profiles, \ie are they really
Sybils? To determine ground-truth, we use the methodology pioneered by Thomas \etal
to locate fake Twitter accounts~\cite{retrospect-imc11}. We monitored the suspicious
Facebook profiles for 6 weeks, and observed 573 became inaccessible. Attempting to
browse these profiles results in the message ``The page you requested was not found,''
indicating that the profile was either removed by Facebook or by the owner. Although
we cannot ascertain the specific reason that these accounts were removed, the use of
widely available photographs as profile images makes it highly likely that these 573
profiles are fakes.

The sole limitation of our Facebook data is that it only includes data from public
profiles. It is unknown if the characteristics of private accounts (legitimate and
Sybil) differ from public ones. This limitation is shared by all studies that rely
on crawled OSN data.

\begin{table}
\begin{small}
\begin{center}
\scalebox{0.9}{
\begin{tabular}{|c|c@{\hspace{0.5em}}c||c|c@{\hspace{0.8em}}c|}
\hline
\multirow{2}{*}{Dataset} & \multicolumn{2}{|c||}{\# of Profiles} & \multirow{2}{*}{Test Group} & \# of & Profiles \\
& Sybil & Legit. & & Testers & per Tester\\
\hline
\multirow{2}{*} {Renren} & \multirow{2}{*} {100} & \multirow{2}{*} {100} & CN Expert & 24 & 100 \\ 
 & & & CN Turker & 418 & 10 \\ 
\hline
\multirow{3}{*} {\tabincell{c}{Facebook \\ US} } & \multirow{3}{*} {32} & \multirow{3}{*} {50} & US Expert & 40 & 50 \\ 
 & & & US Turker & 299 & 12  \\ 
 & & & US Social & 198 & 25  \\ 
\hline
\multirow{2}{*} {\tabincell{c}{Facebook \\ IN} } &\multirow{2}{*} {50} & \multirow{2}{*} {50} & IN Expert & 20 & 100 \\ 
 & & & IN Turker & 342 & 12  \\ 
\hline
\end{tabular}}
\end{center}
\caption{Datasets, test groups, and profiles per tester.}
\label{tab:dataset}
\end{small}
\vspace{-0.1in}
\end{table}

\para{Renren.} We obtained ground-truth data on Sybil and legitimate profiles
on Renren directly from Renren Inc. The security team at Renren gave us
complete information on 1082 banned Sybil profiles, from which we randomly
selected 100 for our user study. Details on how Renren bans Sybil accounts
can be found in~\cite{sybils-imc11}. We collected legitimate Renren profiles using
the same methodology as for Facebook. We seeded a crawler with 4 trustworthy
profiles from people in the lab, crawled 100K friends-of-friends, and then
sampled 100 public profiles. We forwarded these profiles
to Renren's security team and they verified that the profiles belonged
to real users.

\begin{table}
\begin{small}
\begin{center}
\scalebox{0.9}{
\begin{tabular}{|c|c||c|c|c|c|}
\hline
Dataset & Category  & \tabincell{c}{News-\\Feed} & \tabincell{c}{Photos} & \tabincell{c}{Profile\\ Images} & \tabincell{c}{Censored\\ Images} \\
\hline
\multirow{2}{*} {Renren}  & Legit. & 165 & 302 & 10 & 0\\ 
 & Sybil & 30 & 22 & 1.5 & 0.06 \\ 
\hline
\multirow{2}{*} {\tabincell{c}{Facebook\\ US}} & Legit. & 55.62 & 184.78 & 32.86 & 0\\ 
 & Sybil & 60.15 & 10.22 & 4.03 & 1.81 \\ 
\hline
\multirow{2}{*} {\tabincell{c}{Facebook\\ IN}} & Legit. & 55 & 53.37 & 7.27 & 0 \\ 
 & Sybil & 31.6&10.28 &4.44  &  0.08\\ 
\hline
\end{tabular}
}
\end{center}
\caption{Ground-truth data statistics (average number per profile).}
\label{tab:stats}
\end{small}
\vspace{-0.1in}
\end{table}

\para{Summary and Data Sanitization.} Table~\ref{tab:dataset} lists the
final statistics for our three datasets. 
Since the Renren data was provided directly by Renren Inc.,
all profiles are confirmed as either Sybils or legitimate users. For Facebook
US and India, profiles that were banned by Facebook are confirmed Sybils,
and the remaining unconfirmed suspicious profiles are not listed.

During our manual inspection of profiles, we noticed that some
include images of pornography or graphic violence. We determined that it
was not appropriate for us to use these images as part of our user study.
Thus, we manually replaced objectionable images with a grey image containing
the words ``Pornographic or violent image removed.'' This change protects
our test subjects from viewing objectionable images, while still allowing
them to get a sense of the original content that was included in the profile.
Out of 45,096 total images in our dataset, 58 are filtered from Facebook
US, 4 from Facebook India, and 6 from Renren. All objectionable images are
found on Sybil profiles; none are found on legitimate
profiles. 

Finally, we show the basic statistics of ground-truth profiles in Table~\ref{tab:stats}.
Legitimate users have more photo albums and profile photos, while Sybils
have more censored photos. The ``News-Feed'' column shows the average number of items
in the first 5 chronological pages of each user's news-feed. On Facebook, the
news-feed includes many types of items, including wall posts, status updates,
photo tags, \etc. On Renren, the feed {\em only} includes wall posts from friends.

\subsection{Experiment Design}
\compact
Using the datasets in Table~\ref{tab:dataset}, our goal is to
assess the ability of humans to discriminate between Sybil
and legitimate user profiles. To test this, we perform a simple, controlled
study: we show a human test subject (or simply a {\em tester}) a
profile from our dataset, and ask them to classify it as real or fake.
The tester is allowed to view the profile's basic information, wall,
photo albums, and individual photos before making their judgment. If
the tester classifies the profile as fake, they are asked what profile
elements (basic information, wall, or photos) led them to this determination.

Each tester in our study is asked to evaluate several profiles from our dataset,
one at a time. Each tester is given roughly equal number of Sybil
profiles and legitimate profiles. 
The profiles from each group are
randomized for each tester, and the order the profiles are shown in is
also randomized.

\begin{figure}[t]
	\includegraphics[width=0.48\textwidth]{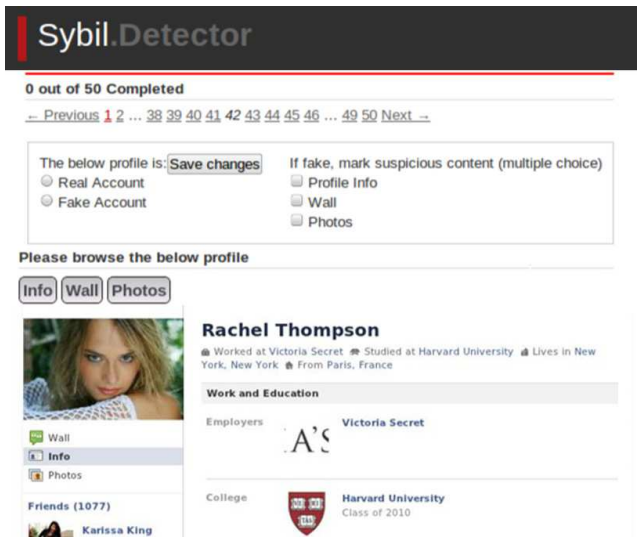}
	\caption{Screenshot of the English version of our user study website.}
	\label{fig:webpage}
	\vspace{-0.1in}
\end{figure}

\para{Implementation.} We implement our study as a website. When a tester
begins the study, they
are presented with a webpage that includes a consent form and details about
our study. After the tester agrees, they are directed to the first profile
for them to evaluate. Figure~\ref{fig:webpage} shows a screenshot of our
evaluation page. At the top are links to the all of the profiles the tester
will evaluate. Testers may use these links to go back and change their
earlier answers if they wish.

Below the numbered links is a box where testers can record their evaluation for
the given profile: real or fake, and if fake, what profile elements are
suspicious (profile, wall, and/or photos)? When the tester is done evaluating
the given profile, they click the ``Save Changes'' button, which automatically
directs their browser to the next profile, or the end of the survey if all
profiles have been evaluated.

Below the evaluation box are three buttons that allow the tester to view the
given profile's basic information (shown by default), wall, and photo albums.
The basic information and wall are presented as JPEG images, in order to preserve the
exact look of Facebook/Renren, while also preventing the tester from clicking
any (potentially malicious) embedded links. Testers may click on each photo
album to view the individual photos contained within.

At the end of the survey, the tester is asked to answer a short questionnaire
of demographic information. Questions include age, gender, country of residence,
education level, and years of OSN experience. There is also a free-form comment
box where tester can leave feedback.

On the server-side, we record all of the classifications and questionnaire
answers made by each tester. We also collect additional information such as
the time spent by the tester on each page, and total session time per tester. 

Because our datasets are in two different languages, we construct two versions
of our study website. Figure~\ref{fig:webpage} shows the English version of our
site, which is used to evaluate Facebook profiles. We also constructed a Chinese
version of our site to evaluate Renren profiles.

\para{Limitations.} The methodology of our user study has two minor
limitations. First, we give testers full profiles to evaluate,
including basic info, wall, and photos. It is not clear how accurate testers
would be if given different information, or a restricted subset of this information.
Second, we assume that there are no malicious testers participating in our
user study. Although attackers might want to infiltrate and disrupt a real
crowdsourced Sybil detector, there is little for them to gain by disrupting
our study. Related work on detecting crowdsourcing abuse may be helpful in
mitigating this problem in the future~\cite{moderate}.

\begin{figure*}[t]
\begin{center}
\mbox{
\hspace{-0.05in}
\subfigure[Education Level]{
   \includegraphics[width=0.335\textwidth] {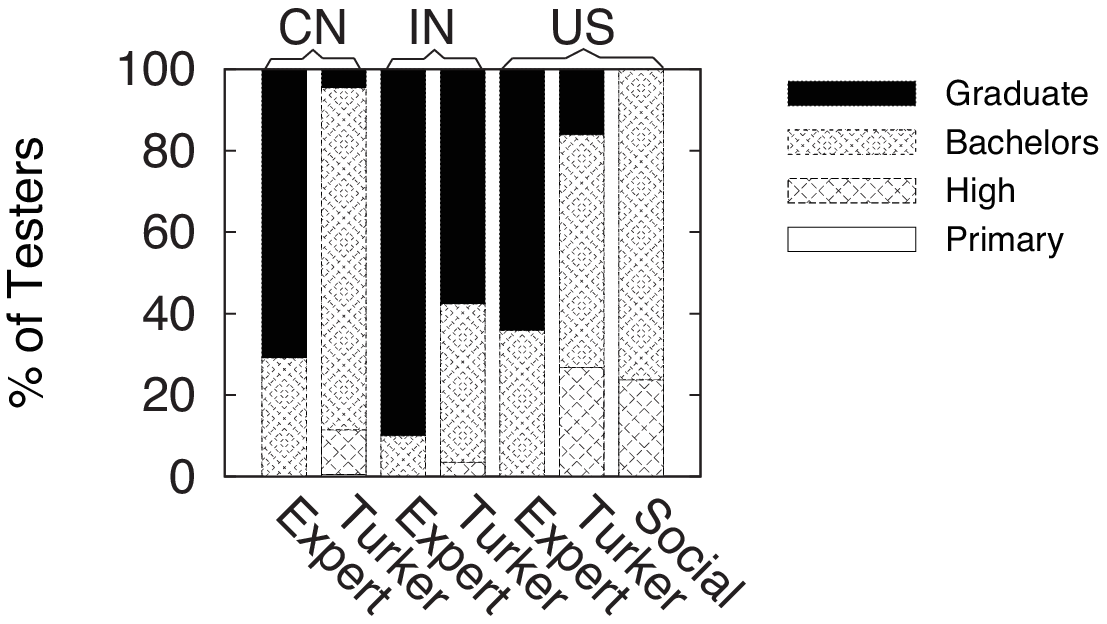}
   \label{fig:workers1}
 }
\hspace{-0.05in}
 \subfigure[OSN Usage Experience]{
   \includegraphics[width=0.335\textwidth] {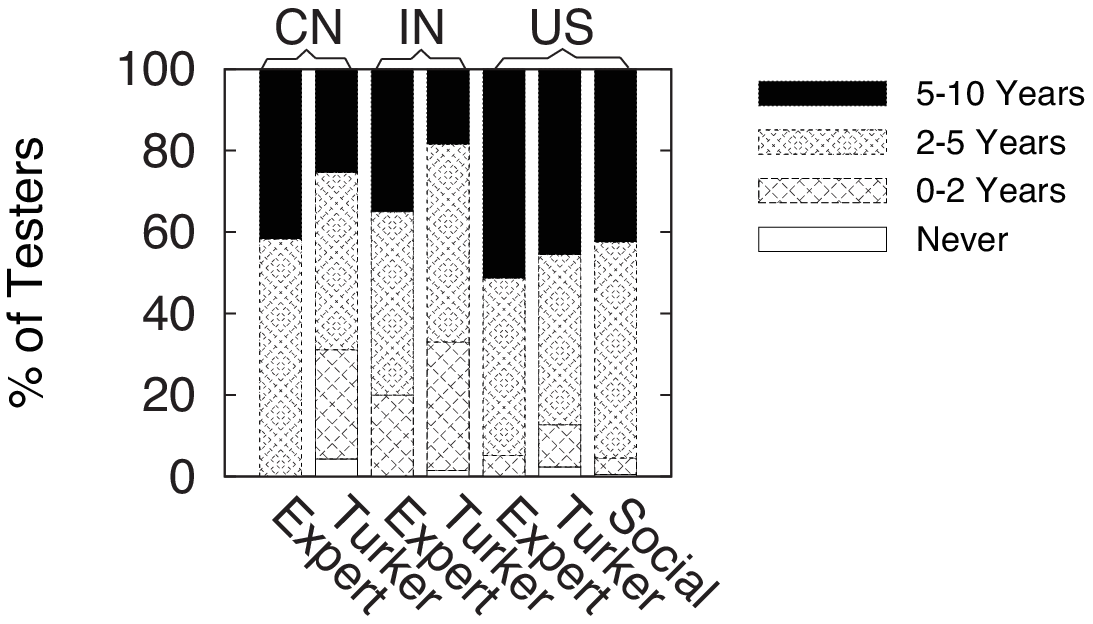}
   \label{fig:workers2}
 }
\hspace{-0.05in}
 \subfigure[Gender]{
   \includegraphics[width=0.335\textwidth] {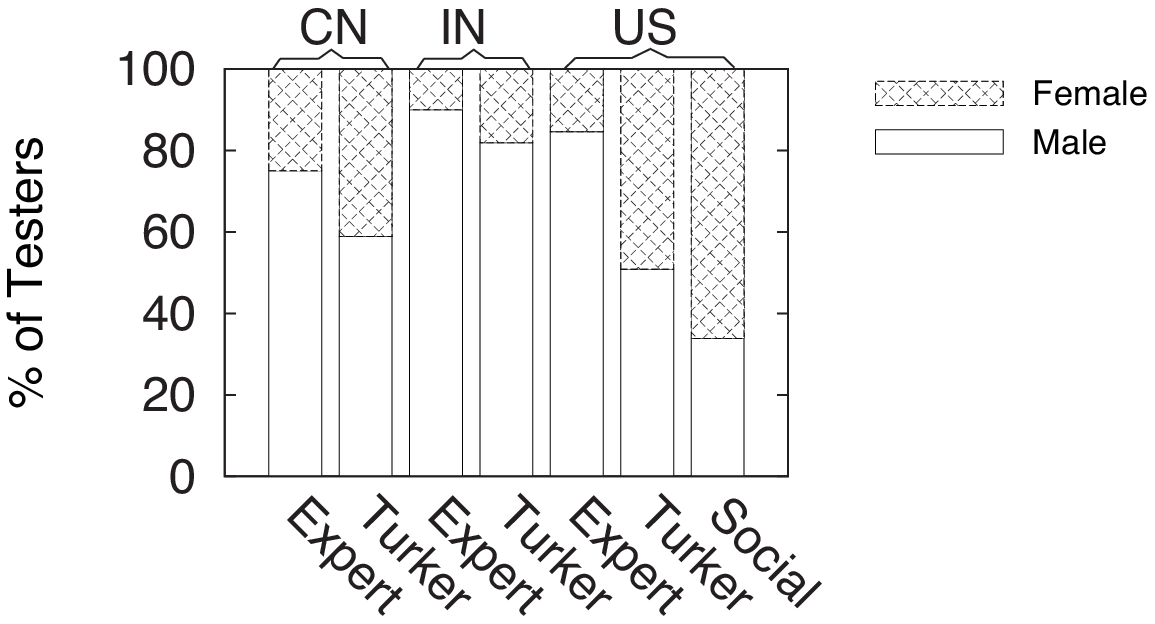}
   \label{fig:workers3}
 }
 }
\vspace{-0.2in}
\caption{Demographics of participants in our user study.}
\label{fig:workers}
\end{center}
\vspace{-0.1in}
\end{figure*}

\subsection{Test Subjects} 
\label{sec:testers}
\compact
In order to thoroughly investigate how accurate different types of users are
at detecting Sybils, we ran user studies on three different groups of test
subjects. Each individual tester was asked to evaluate $\ge$10 profiles
from our dataset, and each profile was evaluated by multiple testers from each
group. This allows us to examine the overall detection accuracy of the group
(\eg the crowd), versus the accuracy of each individual tester. We now
introduce the three test groups, and explain how we administered our study
to them.

\para{Experts.} The first group of test subjects are {\em experts}. This group
contains Computer Science professors and graduate students that were carefully
selected by us. The expert group represents the practical upper-bound on
achievable Sybil detection accuracy.

The expert group is subdivided into three regional groups: US, Indian, and Chinese
experts. Each expert group was evaluated on the corresponding regional dataset.
We approached experts in person, via email, or via social media and directed them
to our study website to take the test. Table~\ref{tab:dataset} lists the number
of expert testers in each regional group. Expert tests were conducted in February, 2012.

As shown in Table~\ref{tab:dataset}, each Chinese and Indian expert evaluated
100 profiles from our dataset, while US experts evaluated 50 profiles. This is
significantly more profiles per tester than we gave to any other test group.
However, since experts are dedicated professionals, we assume that their accuracy
will not be impacted by survey fatigue. We evaluate this assumption in
Section~\ref{sec:aa}.

\para{Turkers.} The second group of test subjects are {\em turkers} recruited from
crowdsourcing websites. Unlike the expert group, the background and education
level of turkers cannot be experimentally controlled. Thus, the detection
accuracy of the turker group provides a lower-bound on the efficacy of a
crowdsourced Sybil detection system.

Like the expert group, the turker group is subdivided into three regional
groups. US and Indian turkers were recruited from MTurk.
HITs on MTurk may have {\em qualifications} associated with them. We used
this feature to ensure that only US based turkers took the Facebook US test, and
Indian turkers took the Facebook India test. We also required that turkers
have $\ge$90\% approval rate for their HITs, to filter out
unreliable workers. We recruited Chinese turkers from {\em Zhubajie}, the
largest crowdsourcing site in China. Table~\ref{tab:dataset} lists
the number of turkers who completed our study in each region. Turker tests
were conducted in February, 2012.

Unlike the expert groups, turkers have an incentive to sacrifice accuracy in favor
of finishing tasks quickly. Because turkers work for pay, the faster they complete
HITs, the more HITs they can do. Thus, of all our test groups, we gave turkers
the fewest number of profiles to evaluate, since turkers are most likely to be
effected by survey fatigue. As shown in Table~\ref{tab:dataset}, Chinese turkers
each evaluated 10 profiles, while US and Indian turkers evaluated 12.

We priced each Zhubajie HIT at \$0.15 (\$0.015 per profile), and each MTurk
HIT at \$0.10 (\$0.0083 per profile). These prices are in line with the
prevailing rates on crowdsourcing websites~\cite{amazonanalysis-acm}.
Although we could have paid more, prior work has shown that paying more
money does not yield higher
quality results on crowdsourcing sites~\cite{financial-hcomp09}.

\para{Sociology Undergraduates.} The final group of test subjects are undergraduate
students from the Department of Communications at UCSB (Social Science
major). These students were asked to take our study in exchange for
course credit. This group adds additional perspective to our study,
apart from Computer Science oriented experts and the uncontrolled
turker population.

The social science students are listed in Table~\ref{tab:dataset} as ``US social.''
We only asked the students to evaluate our Facebook US dataset, since cultural and
language barriers prevent them from effectively evaluating Chinese and Indian
profiles. 198 total students completed our study in March, 2012.
Each student was asked to evaluate 25 profiles, mid-way between what we asked of
experts and turkers.

\para{Summary.} We conduct experiments with 7 groups of testers:
experts from US, India, and China; turkers from US, India, and China, and social science
students from the US. Table~\ref{tab:dataset} lists the number of testers in each group
and the number of profiles evaluated by each tester.

\begin{figure*}[t]
\begin{minipage}{0.31\textwidth}
\begin{center}
	\includegraphics[width=1\textwidth]{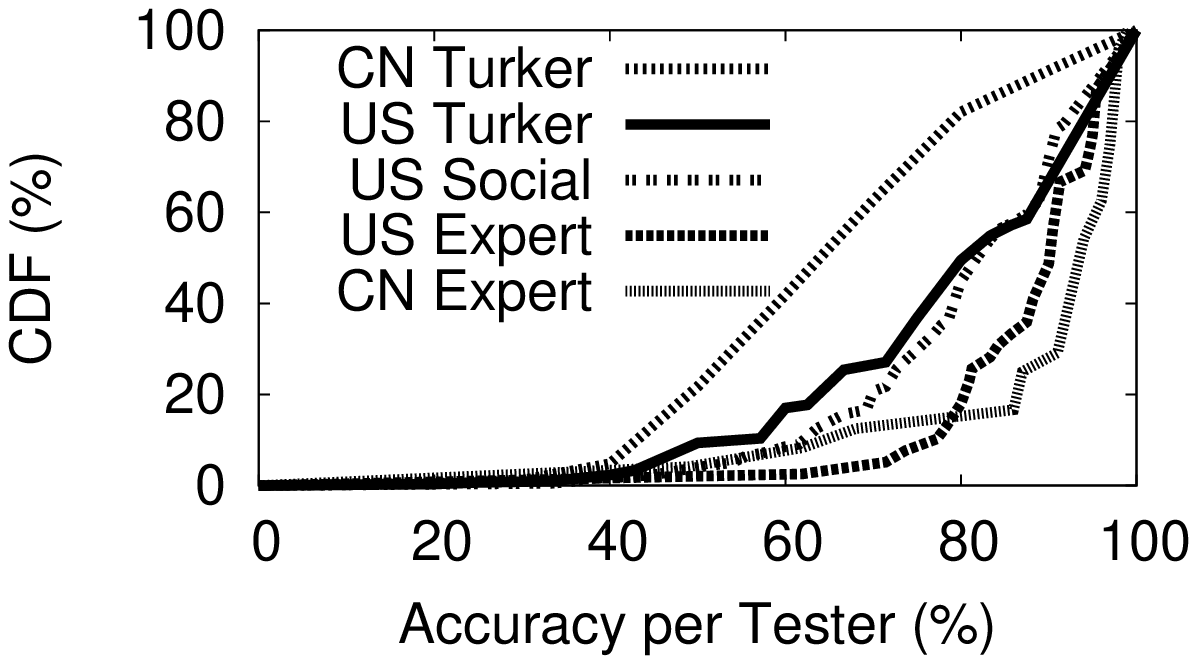}
	\caption{Tester accuracy.}
	\label{fig:worker-acc}
\end{center}
\end{minipage}
\hfill
\begin{minipage}{0.645\textwidth}
\begin{center}
\mbox{
    \subfigure[Renren]{
     	\includegraphics[width=0.5\textwidth]{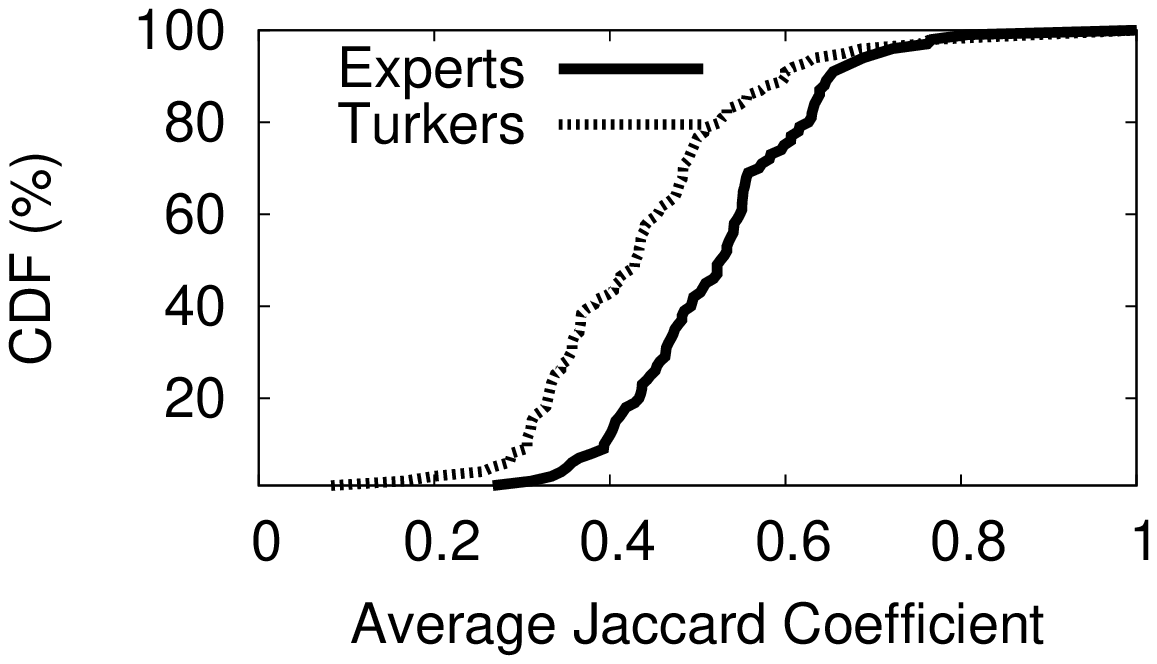}
    	\label{fig:renren-reason}
    }
    \subfigure[Facebook US]{
     	\includegraphics[width=0.5\textwidth]{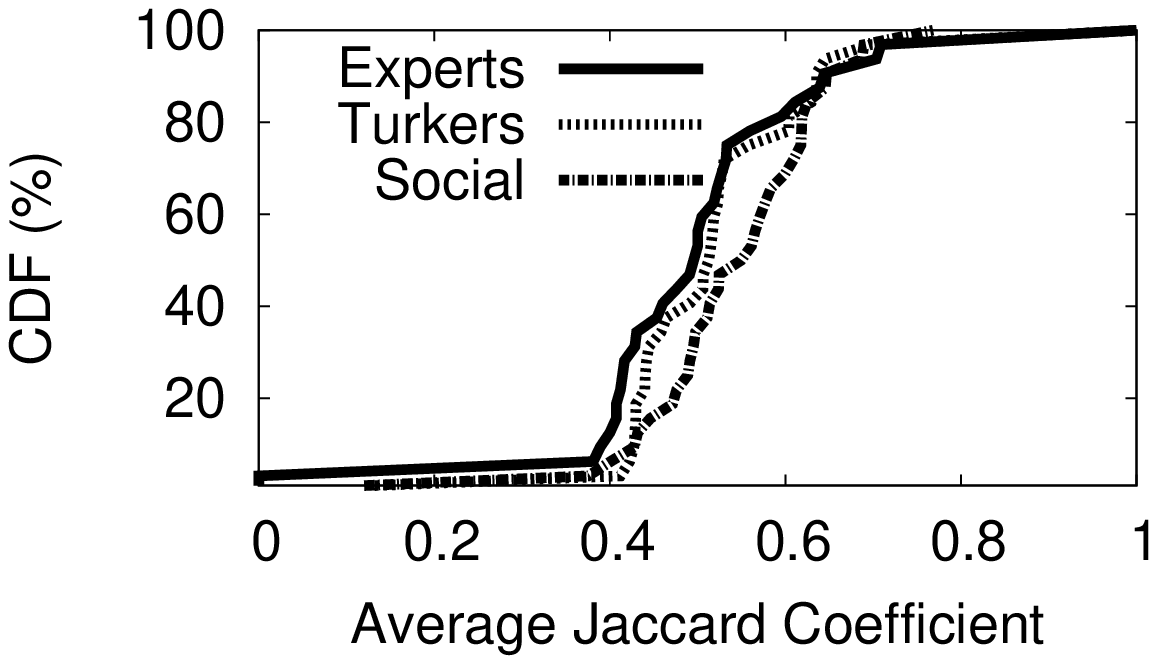}
	    \label{fig:us-reason}
    }
}
\vspace{-0.2in}
\caption{Jaccard similarity coefficient of reasons.}
\label{fig:reason}
\end{center}
\end{minipage}
\end{figure*}

\begin{figure*}[t]
\begin{center}
\mbox{
\subfigure[Renren]{
    \includegraphics[width=0.325\textwidth]{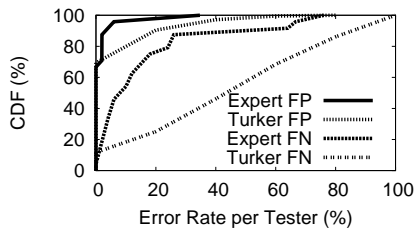}
	\label{fig:worker-fpfn1}
 }
 \subfigure[Facebook US]{
	\includegraphics[width=0.325\textwidth]{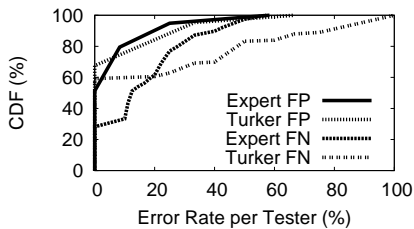}
	\label{fig:worker-fpfn2}
 }
 \subfigure[Facebook India]{
	\includegraphics[width=0.325\textwidth]{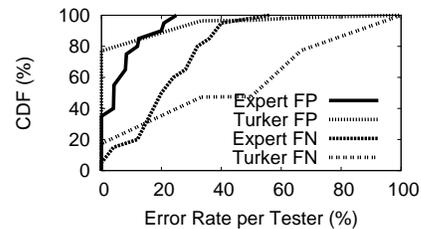}
	\label{fig:worker-fpfn3}
 }
 }
\vspace{-0.2in}
\caption{False positive (FP) and false negative (FN) rates for testers.}
\label{fig:worker-fpfn}
\end{center}
\vspace{-0.1in}
\end{figure*}

\section{User Study Results}
\label{sec:results}
\compact

In this section, we present the high level results of our user study.
We start by introducing the demographics of the test subjects. Next,
we address one of our core questions: how accurate are people
at identifying Sybils? We compare the accuracy of individual testers to
the accuracy of the group to assess whether the ``wisdom of the
crowd'' can overcome individual classification errors. Finally, we examine
the reasons testers cited in classified profiles as Sybils.

\subsection{Demographics} 
\compact
At the end of each survey, testers were asked to answer demographic questions
about themselves. Figure~\ref{fig:workers} shows the results that were
self-reported by testers.

\para{Education.} As shown in Figure~\ref{fig:workers1}, most of our experts
are enrolled in or have received graduate level degrees. This is by design,
since we only asked Computer Science graduate students, undergrads enrolled
in graduate courses, and professors to take part in our expert
experiments. Similarly, the social science testers are drawn from the
undergraduate population at UCSB, which is reflected in the results.

The education levels reported by turkers are surprisingly high. The majority
of turkers in the US and China report enrollment or receipt of
bachelors-level degrees~\cite{demographic-chi10}.  Surprisingly, over 50\% of
Indian turkers report graduate level educations. This result for Indian
turkers stems from cultural differences in how education levels are
denoted. Unlike in the US and China, in India ``graduate school'' refers to
``graduated from college,'' not receipt of a post-graduate degree (\eg
Masters or Doctorate). Thus, most ``graduate'' level turkers in India are
actually bachelors level.

\para{OSN Usage Experience.} As shown in Figure~\ref{fig:workers2}, the vast
majority of testers report extensive experience with OSNs.
US experts, Chinese experts, and social science undergrads almost uniformly
report $\ge$2 years of OSN experience. Indian experts, Indian turkers, and
Chinese turkers have the greatest fractions of users with $<$2 years of OSN
experience. US turkers report levels of OSN experience very similar
to our most experienced expert groups. 

\para{Gender.} As shown in Figure~\ref{fig:workers3}, the vast majority of
our testers are male. 
The only group which exhibits a female majority is the social science
undergrads, a demographic bias of the communications major. 
Turker groups show varying levels of gender bias: Chinese and Indian turkers
are predominantly male~\cite{demographic-chi10}, while the US group is evenly
divided.

\subsection{Individual Accuracy}
\label{sec:indiv_acc}
\compact
We now address one of the core questions of the paper: how accurate are
people at identifying Sybils?  To achieve 100\% accuracy, a tester
needs to correctly classify all Sybil and legitimate profiles they were
shown during the test. 
Figure~\ref{fig:worker-acc} shows the accuracy of testers in 5 of our test
groups. Chinese experts are the most accurate, with half achieving $\ge$90\%
accuracy. The US and Indian (not shown) experts also
achieved high accuracy. However, the turker groups do not perform as well
as the experts. The Chinese and Indian (not shown) turkers perform the worst,
with half achieving $\le$65\% accuracy. The accuracy of US turkers and social
science students falls in-between the other groups.

To better understand tester accuracy, Figure~\ref{fig:worker-fpfn} separates
the results into false positives and false negatives. A false
positive corresponds to misclassifying a legitimate profile as a Sybil, while a false
negative means failing to identify a Sybil. Figure~\ref{fig:worker-fpfn} focuses
on our expert and turker test groups; social science students perform similarly
to US turkers, and the results are omitted for brevity.

Figure~\ref{fig:worker-fpfn} reveals similar trends across all test groups. First,
false positives are uniformly lower than false negatives, \ie testers
are more likely to misclassify Sybils as legitimate, than vice versa. Second,
in absolute terms, the false positive rates are quite low: $<$20\% for 90\% of
testers. Finally, as in Figure~\ref{fig:worker-acc}, error rates for turkers tend
to be significantly higher than those of experts.

In summary, our results reveal
that people can identify differences between Sybil and legitimate
profiles, but most individual testers are not accurate enough to
be reliable.  

\subsection{Accuracy of the Crowd}
\label{sec:crowdaccuracy}
\compact
We can leverage ``the wisdom of the crowd'' to
amortize away errors made by individuals. Many studies on crowdsourcing
have demonstrated that experimental error can be controlled by having multiple turkers
vote on the answer, and then using the majority opinion as the final
answer~\cite{snow-emnlp08, Le-SIGIR10}. As long as errors by turkers
are uncorrelated, this approach generates very accurate results.

We now examine whether this methodology can be used to improve the classification
accuracy of our results. This question is of vital importance,
since a voting scheme would be an essential component of a crowdsourced Sybil
detector. To compute the ``final'' classification for each profile in our dataset,
we aggregate all the votes for that profile by testers in each group.
If $\ge$50\% of the testers vote for fake, then we classify that profile as a Sybil.

Table~\ref{tab:results} shows the percentage of false positive and negative
classifications for each test group after we aggregate votes. The results
are mixed: on one hand, false positive rates are uniformly low across all
test groups.  In the worst case, US turkers and social science students only
misclassify 1 out of 50 legitimate profiles.  Practically, this means
that crowds can successfully identify real OSN profiles.

On the other hand, false negative rates vary widely across test groups.
Experts in China, in the US, and the social science students
all perform well, with false negative rates $<$10\%.  Indian experts also
outperform the turker groups, but only by a 2.7\% margin. The Chinese
and Indian turker groups perform worst, with $\ge$50\% false negatives.

\begin{table}
\begin{small}
\begin{center}
\begin{tabular}{|c|c|c c|}
\hline
Dataset & Tester  & FP Rate & FN Rate  \\
\hline
\multirow{2}{*} {Renren}  & CN Expert & 0\% & 3\% \\ 
 & CN Turker & 0\% & 63\% \\ 
\hline
\multirow{3}{*} {Facebook US } & US Expert & 0\% & 9.4\% \\ 
 & US Turker & 2\% & 18.7\% \\ 
 & US Social & 2\% & 6.25\% \\ 
\hline
\multirow{2}{*} {Facebook IN} & IN Expert & 0\% & 16\%\\ 
 & IN Turker & 0\%  & 50\% \\ 
\hline
\end{tabular}
\caption{Error rates after aggregating votes.}
\label{tab:results}
\end{center}
\end{small}
\vspace{-0.1in}
\end{table}

From these results, we can conclude three things.  First, using aggregate
votes to classify Sybils {\em does} improve overall accuracy significantly. Compared to
the results for individual testers in Figure~\ref{fig:worker-fpfn}, both
false positive and negative rates are much lower after aggregation. Second,
the uniformly low false positive rates are a very good result. This means
that running a crowdsourced Sybil detection system will not 
harm legitimate social network users. Finally, even with
aggregate voting, 
turkers are still not as accurate as experts. In the next section, we
look more deeply into factors that may negatively influence turkers accuracy,
and techniques that can mitigate these issues.

\begin{figure*}[t]
\begin{center}
\mbox{
\subfigure[US Turker]{
    \includegraphics[width=0.3\textwidth]{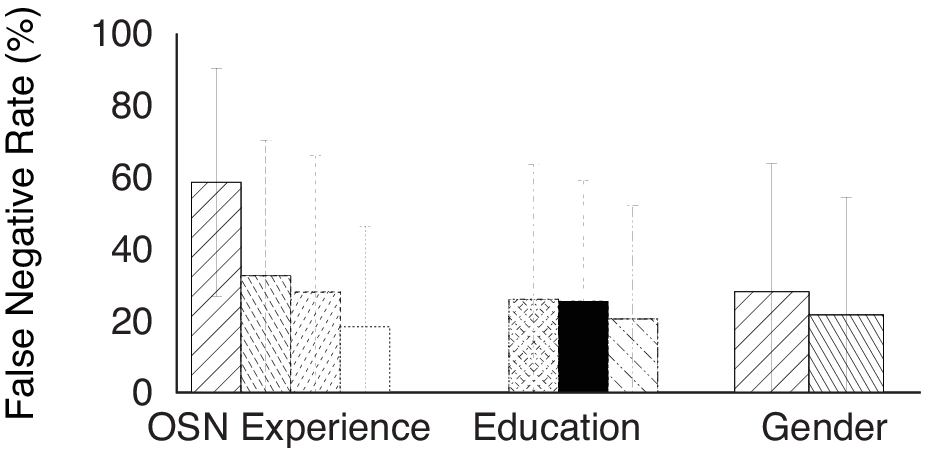}
	\label{fig:worker-break1}
 }
 \hspace{-0.1in}
 \subfigure[India Turker]{
	\includegraphics[width=0.3\textwidth]{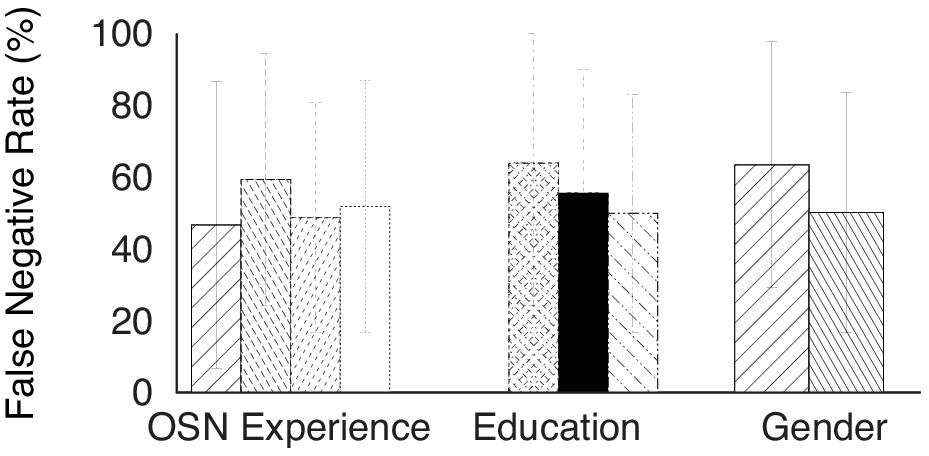}
	\label{fig:worker-break2}
 }
 \hspace{-0.1in}
 \subfigure[Chinese Turker]{
	\includegraphics[width=0.39\textwidth]{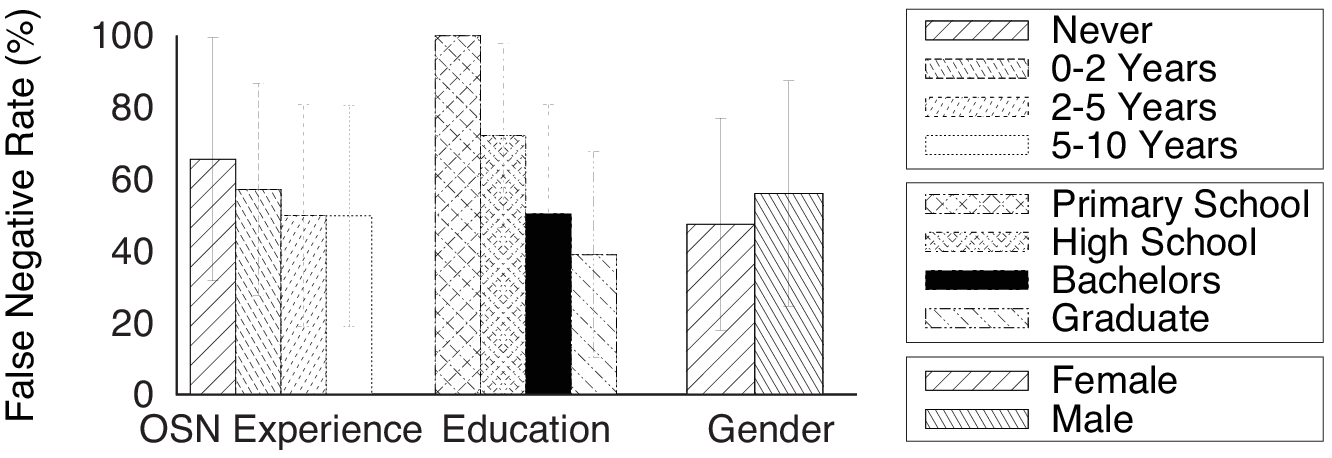}
	\label{fig:worker-break3}
 }
}
\vspace{-0.2in}
\caption{False positive rates for turkers, broken down by demographic.}
\label{fig:worker-break}
\end{center}
\vspace{-0.1in}
\end{figure*}

\begin{table}
\begin{small}
\begin{center}
\begin{tabular}{|c|c|c c c |}
\hline
Dataset & Tester  & Info & Wall & Photos  \\
\hline
\multirow{2}{*} {Renren}  & CN Expert & 18\% & 57\% & 25\%\\ 
 & CN Turker & 31\% & 31\% & 38\% \\ 
\hline
\multirow{3}{*} {Facebook US} & US Expert & 37\% & 30\% & 33\% \\ 
 & US Turker & 35\% & 32\% & 33\% \\ 
 & US Social & 30\% & 31\%  & 39\% \\ 
\hline
\multirow{2}{*} {Facebook IN} & IN Expert & 39\% & 28\% & 33\%\\ 
 & IN Turker & 39\% & 27\%  & 34\% \\ 
\hline
\end{tabular}
\caption{Reasons why profiles are suspicious.}
\label{tab:reasons}
\end{center}
\end{small}
\vspace{-0.1in}
\end{table}

\subsection{Reasons for Suspicion}
\compact
During our user study, testers were asked to give {\em reasons} for
why they classified profiles as Sybils. Testers were given the option
of reporting the profile's basic information, wall, and/or photos as
suspicious. Testers could select as many options as they liked.

In this section, we compare and contrast the reasons reported by different
test groups.  Table~\ref{tab:reasons} shows percentage of votes for each
reasons across our seven test groups. The US and Indian expert and turker
groups are very consistent: they all slightly favor basic information. The
bias may be due to the way our study presented information, since each
profile's basic information was shown first, by default. The social science
students are the only group that slightly favors photos.

In contrast to the US and Indian groups, Chinese experts and turkers often disagree
on their reasons for suspicion. The majority of experts rely on wall messages, while
turkers slightly favor photos.  As shown in Figure~\ref{fig:worker-acc}, Chinese turkers
have lower accuracy than Chinese experts.  One possible reason for this result
is that turkers did not pay enough attention to the wall.  As previously mentioned,
there is a comment box at the end of our survey for testers to offer feedback and
suggestions. Several Chinese experts left comments saying they observed
wall messages asking questions like ``do I know you?,'' and ``why did you send me a
friend request?,'' which they relied on to identify Sybil profiles.

\para{Consistency of Reasons.} There is no way to objectively evaluate
the correctness of tester's reasons for classification, since there is
no algorithm that can pick out suspicious pieces of information from an OSN profile.
Instead, what we can do is examine how consistent the reasons are for each
profile across our test groups. If all the testers agree on the reasons
why a given profile is suspicious, then that is a strong indication that
those reasons are correct.

To calculate consistency, we use the following procedure. In each test group,
each Sybil is classified by $N$ testers.  For all pairs of users in each
group that classified a particular Sybil profile, we calculate the Jaccard
similarity coefficient to look at overlap in their reasons, giving us
$N*(N-1)/2$ unique coefficients. We then compute the average of these
coefficients for each profile. By computing the average Jaccard coefficient
for each Sybil, we arrive at a distribution of consistency scores for all
Sybils for a given test group.

Figure~\ref{fig:reason} shows the consistency distributions of the China and US
test groups. The results for the Indian test groups are similar to US testers,
and are omitted for brevity. The Chinese turkers show the most disagreement:
for 50\% of Sybils the average Jaccard coefficient is
$\le$0.4. Chinese experts and all three US groups exhibit similar levels of
agreement: 50\% of Sybils have coefficients $\le$0.5. The fraction of Sybils
receiving near complete disagreement (0.0) or agreement (1.0) is negligibly
small across all test groups.

Based on these results, we conclude that testers identify Sybils for
inconsistent reasons. Even though Table~\ref{tab:reasons} shows that
each of the three available reasons receives a roughly equal portion
of votes overall, the reasons are assigned randomly across Sybils
in our dataset. This indicates that no single profile feature is a
consistent indicator of Sybil activity, and that testers benefit from
having a large, diverse set of information when making classifications.  Note
that this provides further support that automated mechanisms based on
individual features are less likely to succeed, and also explains the
success of human subjects in detecting Sybils.

\para{Answer Revisions.} While taking our survey, testers had the option of
going back and changing
classifications that they have previously made. However, few took advantage
of this feature. This is not unexpected, especially for turkers.
Since turkers earn more if they work faster, there is a negative incentive
to go back and revise work. In total, there were only 28 revisions by testers:
16 from incorrect to correct, and 12 from correct to incorrect.

\begin{figure*}[t]
\begin{center}
\mbox{
 \subfigure[Renren Expert]{
	\includegraphics[width=0.24\textwidth]{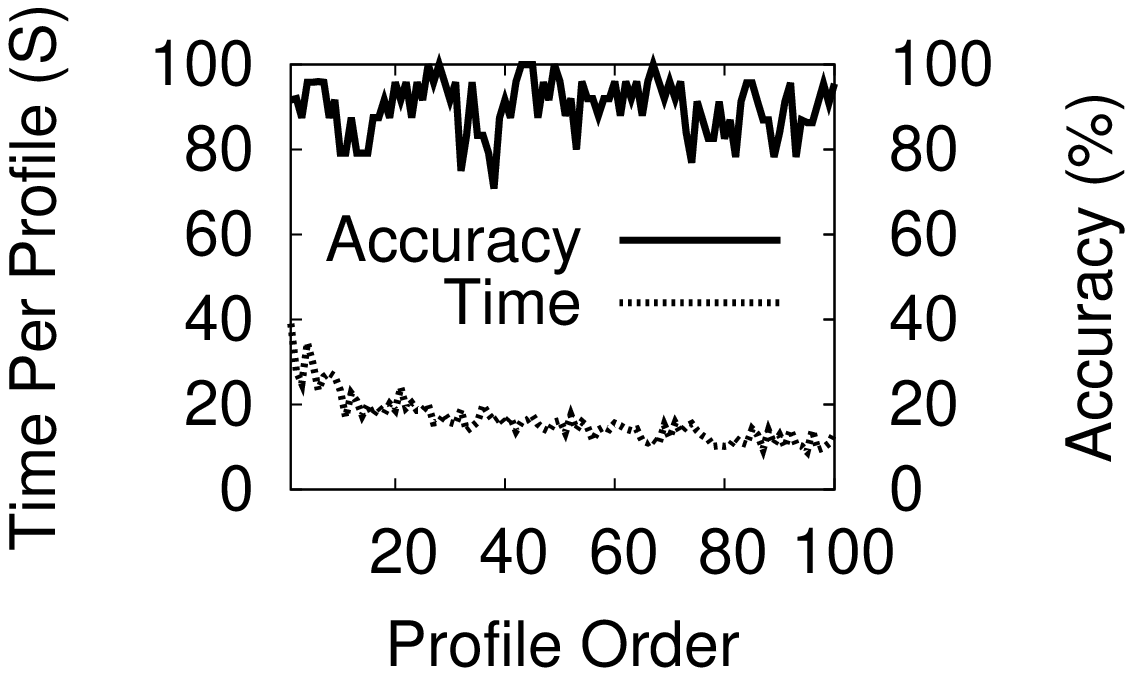}
	\label{fig:renren-pro-acctime}
 }
 \subfigure[Renren Turker]{
	\includegraphics[width=0.24\textwidth]{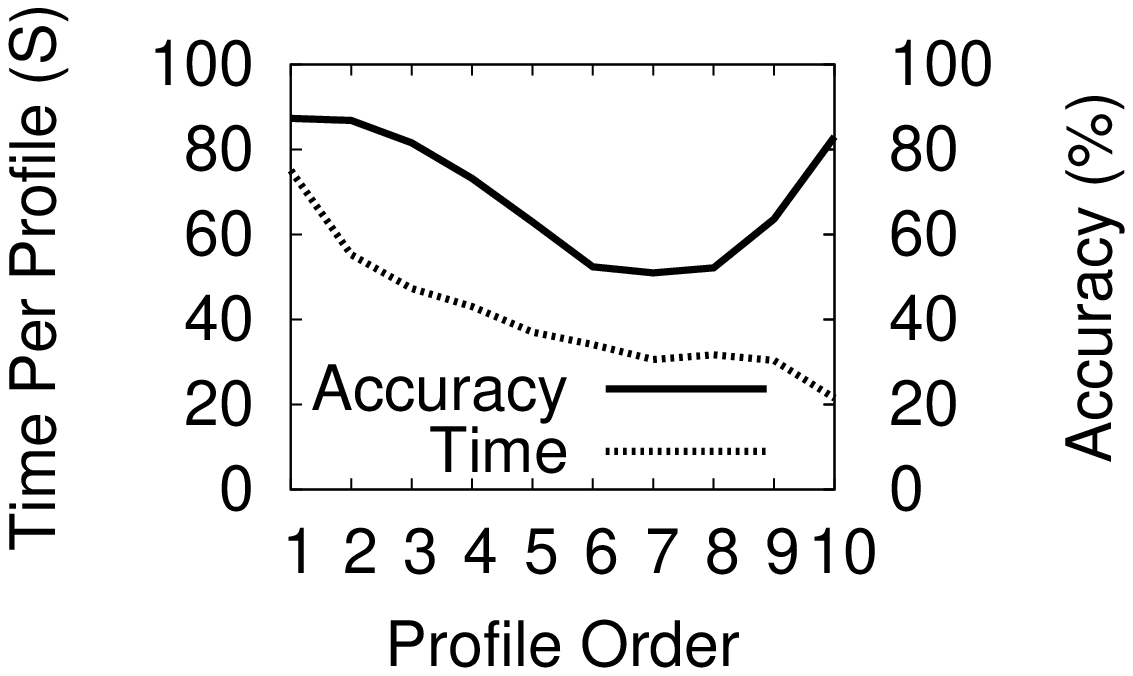}
	\label{fig:renren-turk-acctime}
 }
 \subfigure[Facebook US Expert]{
	\includegraphics[width=0.24\textwidth]{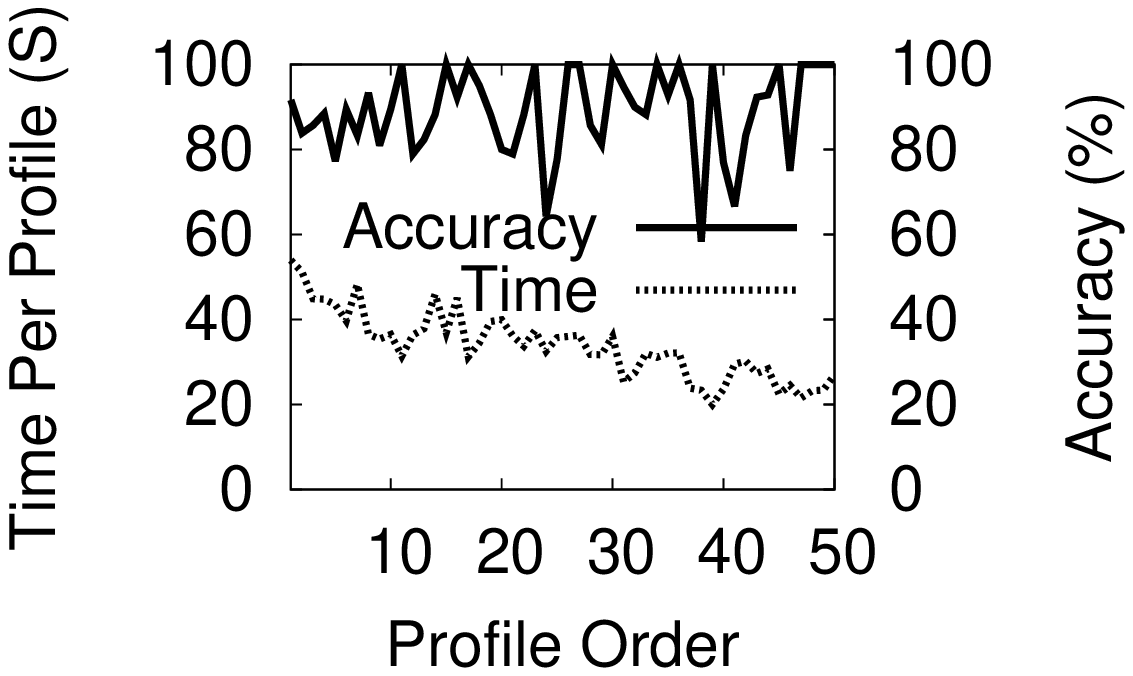}
	\label{fig:us-pro-acctime}
 }
 \subfigure[Facebook US Turker]{
	\includegraphics[width=0.24\textwidth]{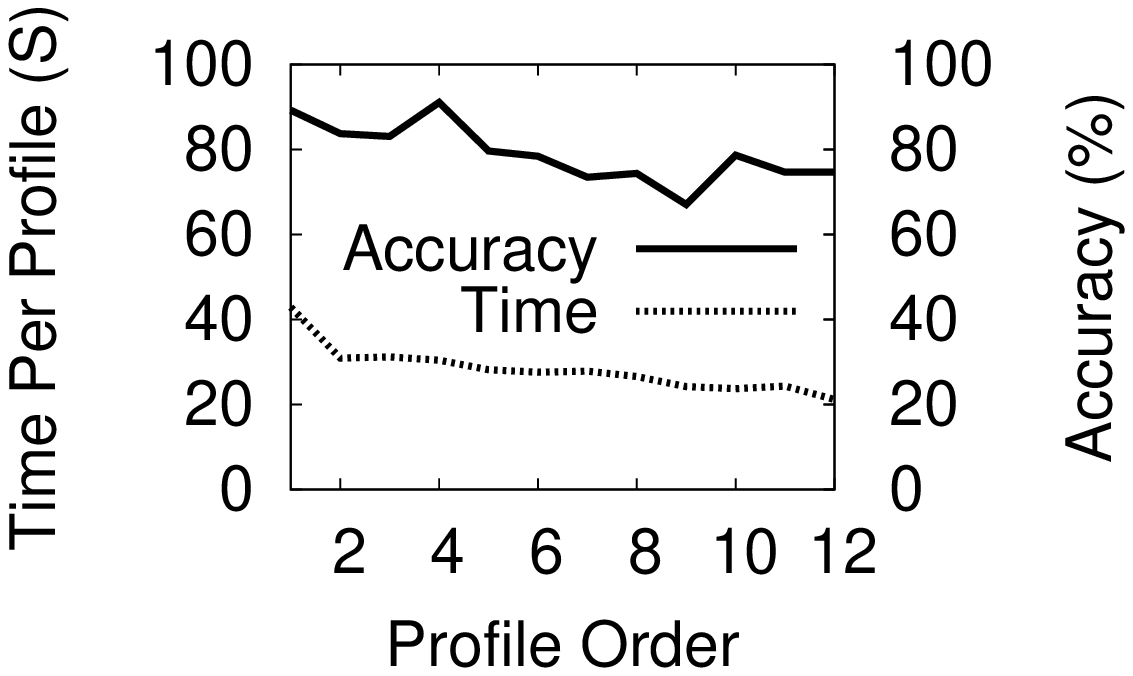}
	\label{fig:us-turk-acctime}
 }
}
\vspace{-0.2in}
\caption{Accuracy and time on the $n$th profile.}
\label{fig:acc-time}
\end{center}
\vspace{-0.1in}
\end{figure*}

\section{Turker Accuracy Analysis}
\label{sec:aa}

The end goal of our work is to create a crowdsourced Sybil detection
system. However, in Section~\ref{sec:results} we observed that turkers are
not as accurate as experts. In this section, we examine factors that may
impact the accuracy of turkers, and investigate ways to improve our Sybil
detection system. We start by looking at demographic factors. Next, we
examine profile evaluation time to understand if turkers are adversely
affected by survey fatigue. Next, we examine issues of turker selection.
Will adding more turkers to the crowd improve accuracy? What if we set a
threshold and filter out turkers that consistently perform poorly? Finally, we
calculate the per profile accuracy of testers to detect ``stealth Sybils''
that are undetectable by both experts and turkers.

\subsection{Demographic Factors}
\compact 

First, we explore the impact of demographic factors on the turker's
accuracy. We focus on false negative rates of turkers, since their false
positive rates are close to zero.  Figure~\ref{fig:worker-break} shows the
average false negative rate and standard deviation of turkers from China, US
and India, broken down by different demographics. Education has a clear
impact on false negatives: higher education level correlates with increased ability
to identify Sybils. The impact of OSN usage experience is less clear. Chinese
and US turker's false negative rates decline as OSN experience increases,
which is  expected.  However, for Indian turkers there is no
correlation. Gender does not appear to impact false negatives in a meaningful
way.  The results in Figure~\ref{fig:worker-break} indicate that turker
accuracy could be improved by filtering out workers with few years of OSN
experience and low education level.

\subsection{Temporal Factors and Survey Fatigue} 
\compact 

It is known that turkers try to finish tasks as quickly as possible in order
to earn more money in a limited amount of time~\cite{userstudies-chi08}.
This leads to our next question: do turkers spend less time evaluating
profiles than experts, and does this lead to lower accuracy?  The issue of
time is also related to survey fatigue: does the accuracy of each tester
decrease over time due to fatigue and boredom?

To understand these temporal factors, we plot Figure~\ref{fig:acc-time},
which shows the average evaluation time and accuracy per profile ``slot''
for Chinese and US experts and turkers. The x-axis of each subfigure denotes
the logical order in which testers evaluated profiles, \eg ``Profile Order''
$n$ is the $n^{th}$ profile evaluated by each tester. Note that profiles are
presented to each tester in random order, so each tester evaluated a
different profile within each slot. Within each slot, we calculate the average
profile evaluation time and accuracy across all testers. 100\% accuracy
corresponds to all testers correctly classifying the $n$th profile they were
shown.  Although experts evaluated $>$10 profiles each, we only show the first
10 to present a fair comparison versus the turkers. The results for the Indian
test groups are similar to the US groups, and are omitted for brevity.

The first important result from Figure~\ref{fig:acc-time} is that
absolute profile evaluation time is not a good indicator of accuracy.
The Chinese experts exhibit the fastest evaluation times, averaging
one profile every 23 seconds. However, they are more accurate
than Chinese turkers who spend more time on each profile. This pattern
is reversed on Facebook: experts spend more time and are more
accurate than turkers.

Next, we look for indications of survey fatigue. In all 4 subfigures
of Figure~\ref{fig:acc-time}, the evaluation time per profile decreases
over time. This shows that testers speed up as they progress through the
survey. As shown in the expert Figures~\ref{fig:renren-pro-acctime}
and~\ref{fig:us-pro-acctime}, this speedup does not affect accuracy.
These trends continue through the evaluation of additional profiles (10-50
for Chinese experts, 10-100 for US experts) that are not shown here.
However, for turkers, accuracy does tend to decrease over time, as shown
in Figures~\ref{fig:renren-turk-acctime} and~\ref{fig:us-turk-acctime}. This
demonstrates that turkers are subject to survey fatigue. The up-tick in
Chinese turker accuracy around profile 10 is a statistical anomaly, and is
not significant.

\subsection{Turker Selection}
\compact
As demonstrated in Section~\ref{sec:crowdaccuracy}, we can mitigate the
classification errors of individuals by aggregating their votes together.
This raises our next question: can we continue to improve the overall
accuracy of turkers by simply adding more of them?

To evaluate this, we conducted simulations using the data from our
user study. Let $C$ be the list of classifications received by a given
profile in our dataset (either a Sybil or legitimate profile) by a given
group of turkers (China, US, or India). To conduct our
simulation, we randomize the order of $C$, then calculate what the
overall false positive and negative rates would be as we
include progressively more votes from the list. For each profile, we
randomize the list and conduct the simulation 100 times, then average
the rates for each number of votes. Intuitively, what this process
reveals is how the accuracy of the turker group changes as we increase
the number of votes, irrespective of the specific order that the votes
arrive in.

\begin{figure*}[t]
\begin{minipage}{0.48\textwidth}
\begin{center}
     	\includegraphics[width=0.7\textwidth]{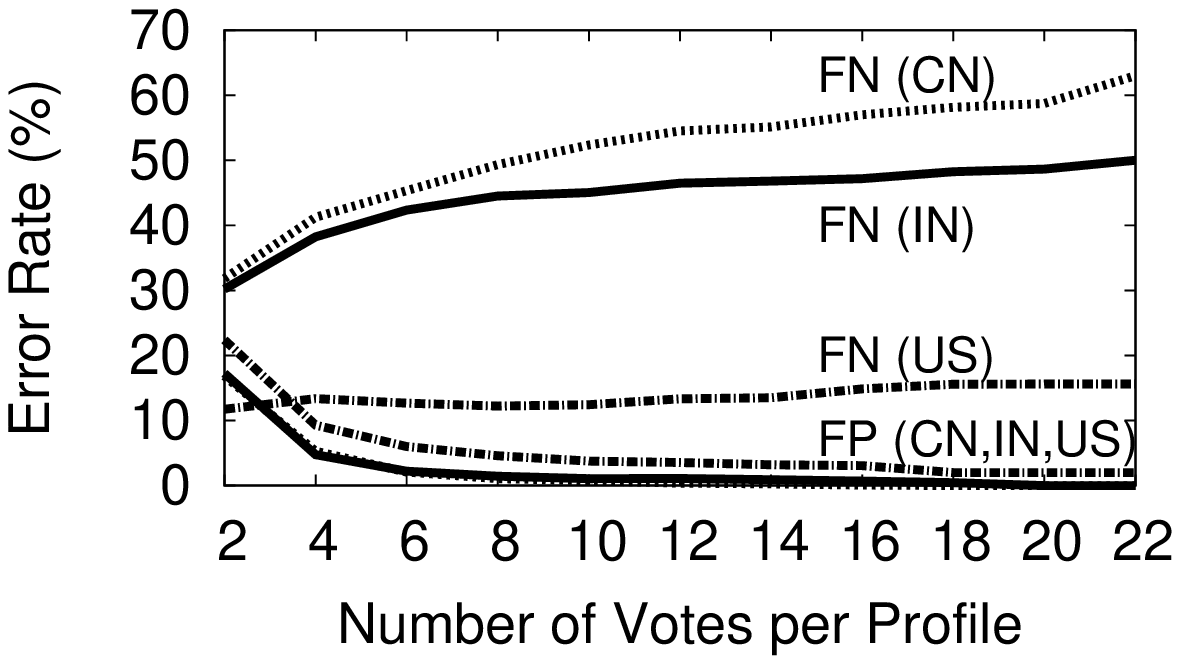}
	\caption{Votes per profile versus the FP and FN rate. }
	\label{fig:cn-wnum-fpfn}
\end{center}
\end{minipage}
\hfill
\begin{minipage}{0.48\textwidth}
\begin{center}
	\includegraphics[width=0.7\textwidth]{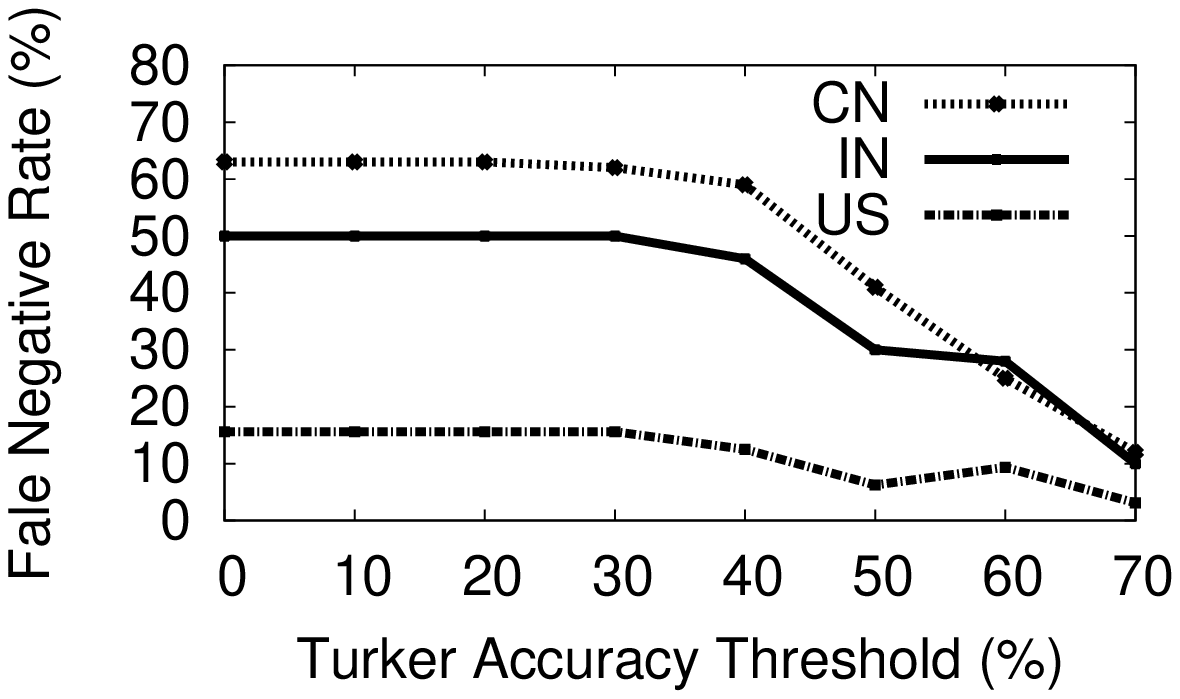}
	\caption{Accuracy threshold versus the false negatives.}
	\label{fig:turker-thre}
\end{center}
\end{minipage}
\hfill
\vspace{-0.1in}
\end{figure*}

The results of our simulations demonstrate that there are limits
to how much accuracy can be improved by adding more turkers
to the group, as shown in Figure~\ref{fig:cn-wnum-fpfn}.
Each line plots the average accuracy over all Sybil and legitimate profiles
for a given group of turkers. For false positives, the trend is very clear:
after 4 votes, there are diminishing returns on additional
votes. For false negatives, the trend is either flat (US turkers), or
it grows slightly worse with more votes (China and India).


\begin{figure*}[t]
\begin{center}
\mbox{
\subfigure[Renren]{
     	\includegraphics[width=0.32\textwidth]{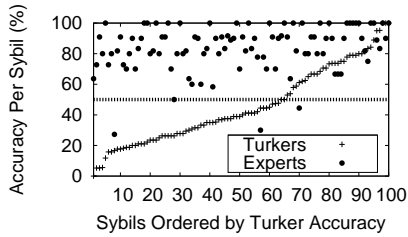}
	\label{fig:renren-sid-diff}
 }
 \subfigure[Facebook US]{
     	\includegraphics[width=0.32\textwidth]{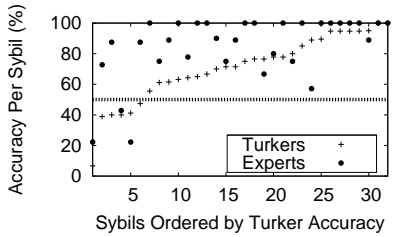}
	\label{fig:us-sid-diff}
 }
 \subfigure[Facebook India]{
     	\includegraphics[width=0.32\textwidth]{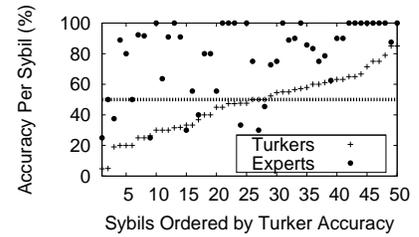}
	\label{fig:india-sid-diff}
 }
 }
\vspace{-0.2in}
\caption{Scatter plots of average accuracy per Sybil profile.}
\label{fig:sid-diff}
\end{center}
\vspace{-0.1in}
\end{figure*}

\para{Filtering Inaccurate Turkers.} Since adding more turkers does not
significantly increase accuracy, we now investigate the opposite approach:
eliminating turkers that are consistently inaccurate. Many deployed
crowdsourcing systems already use this approach~\cite{gold-HCI11}. Turkers
are first asked to complete a pre-screening test, and only those who perform
sufficiently well are allowed to work on the actual job.

In our scenario, turkers could be pre-screened by asking them to classify
accounts from our ground-truth datasets. Only those that correctly classify
$x$ accounts, where $x$ is some configurable threshold, would be permitted
to work on actual jobs classifying suspicious accounts.

To gauge whether this approach improves Sybil detection accuracy, we
conduct another simulation.  We vary the accuracy threshold $x$, and
at each level we select all turkers that have overall accuracy $\ge x$.
We then plot the false negative rate of the selected turkers in
Figure~\ref{fig:turker-thre}.  Intuitively, this simulates turkers
taking two surveys: one to pre-screen them for high accuracy, and a
second where they classify unknown, suspicious accounts.

Figure~\ref{fig:turker-thre} demonstrates that the false negative
rate of the turker group can be reduced to the same level as experts
by eliminating inaccurate turkers. The false negative rates are stable
until the threshold grows $>$40\% because, as shown in Figure~\ref{fig:worker-acc},
almost all the turkers have accuracy $>$40\%. By 70\% threshold, all
three test groups have false negative rates $\le$10\%, which is on par
with experts.  We do not increase the threshold beyond 70\% because it
leaves too few turkers to cover all the Sybil profiles in our dataset.
At the 70\% threshold, there are 156 Chinese, 137 Indian, and 223
US turkers available for work.

\subsection{Profile Difficulty}
\compact
The last question we examine in this section is the following: are there extremely
difficult ``stealth'' Sybils that resist classification by both turkers and experts? As we 
show in Table~\ref{tab:results}, neither experts nor turkers have 0\% false negatives
when classifying Sybils. What is unknown is if there is correlation
between the false negatives of the two groups.

To answer this question, we plot Figure~\ref{fig:sid-diff}. Each scatter plot
shows the average classification accuracy of the Sybils from a particular region.
The x-axes are presented in ascending order by turker accuracy.
This is why the points for the turkers in each subfigure appear to form a line.

Figure~\ref{fig:sid-diff} reveals that, in general, experts can correctly classify the
vast majority of Sybils that turkers cannot (\eg turker accuracy $<$50\%). There are a
select few, extremely difficult
Sybils that evade both the turkers and experts. These ``stealth'' Sybils represent the
pinnacle of difficulty, and blur the line between real and fake user profiles. There is
only one case, shown in Figure~\ref{fig:renren-sid-diff}, where turkers correctly identify
a Sybil that the experts missed.

One important takeaway from Figure~\ref{fig:sid-diff} is that ``stealth'' Sybils are a
very rare phenomenon. Even if a crowdsourced Sybil detector was unable to identify the
them, the overall detection accuracy is so high that most Sybils will be caught
and banned. This attrition will drive up costs for attackers, deterring
future Sybil attacks.

\para{Turker Accuracy and Luck.} Another takeaway from Figure~\ref{fig:sid-diff} is
that some profiles are difficult for turkers to classify.  This leads to a
new question: are the most accurate turkers actually better workers, or
were they just lucky during the survey?  Hypothetically, if a turker was randomly
shown all ``easy'' Sybils, then they would appear to be accurate,
when in fact they were just lucky.

Close examination of our survey results reveals that accurate turkers were not lucky.
The 75 Chinese turkers who achieved $\ge$90\% accuracy were collectively shown 97\% of Renren Sybils
during the survey. Similarly, the 124 US turkers with $\ge$90\% accuracy were also
shown 97\% of the Facebook US Sybils.  Thus, the high accuracy turkers exhibit
almost complete coverage of the Sybils in our dataset, not just the ``easy'' ones.

\section{A Practical System}
\label{sec:prac}
\compact

In this section, we design a crowdsourced Sybil detection system based
on the lessons learned from our experiments. We focus on practical issues
such as scalability, accuracy, and privacy. We first describe our system
architecture that enables crowdsourced Sybil detection at large
scale. Second, we use trace-driven simulations to examine the tradeoff
between accuracy and cost in such a system.  Finally, we discuss how to
preserve user privacy when distributing user profile data to turkers.

\subsection{System Design and Scalability}
\begin{figure}[t]
\centering
     	\includegraphics[width=0.47\textwidth]{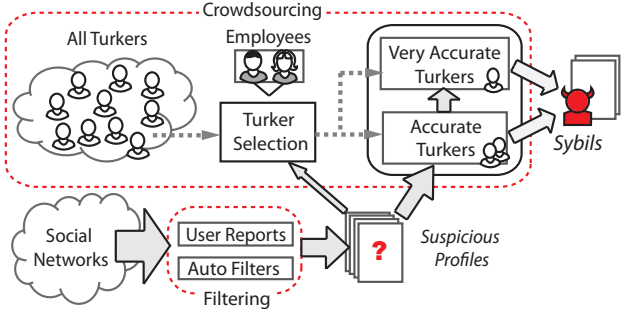}
	\caption{Crowdsourced Sybil detector.}
	\label{fig:detector}
	\vspace{-0.1in}
\end{figure}

\compact
The first challenge we address is scalability. Today's social networks include hundreds
of millions of users, most of whom are legitimate.  How do we build a
system that can focus the efforts of turkers on the subset of accounts that
are suspicious?  To address this challenge, we propose a hierarchical Sybil
detection system that leverages both automated techniques and crowdsourced
human intelligence.  As shown in Figure~\ref{fig:detector}, the system
contains two main layers: the \emph{filtering layer} and the
\emph{crowdsourcing layer}.

\para{Filtering Layer.} In the first layer, we use an ensemble of filters to
locate suspicious profiles in the social network. These filters
can be automated using techniques from prior work, such as Sybil community
detection~\cite{sybilrank} and feature based selection~\cite{sybils-imc11}.
Filters can also be based on existing ``user report'' systems that allow OSN
users to ``report'' or ``flag'' suspicious profiles.  These tools are already
implemented in social networks such as Facebook and Renren, and help to dramatically
reduce the number of target profiles studied by our crowdsourcing layer.

\para{Crowdsourcing Layer.} The output of the filtering layer is a set of
suspicious profiles that require further validation
(Figure~\ref{fig:detector}). These profiles are taken as input by the
crowdsourcing layer, where a group of turkers classify them as legitimate or
fake.  OSNs can take further action by either banning fake accounts, or
using additional CAPTCHAs to limit potentially damaging behavior.

We begin with two techniques to increase the accuracy of the turker
group. First, we use majority voting by a group of workers to classify each
suspicious profile.  Our earlier results in Table~\ref{tab:results} show
that the accuracy of the crowd is significantly better than the accuracy of
individual turkers.

The second mechanism is a ``turker selection'' module that filters out
inaccurate turkers.  Figure~\ref{fig:turker-thre} shows that
eliminating inaccurate turkers drastically reduces the false negative
rate.  As shown in Figure~\ref{fig:detector}, we can implement turker
selection by randomly mixing in ``ground-truth profiles'' that have been
verified by social network exployees with the larger set of suspicious
profiles.  By
examining a tester's answers on the ground-truth profiles, we can gauge
the evaluation accuracy of each worker. This accuracy test is performed
continuously over time, so that any significant deviation
in the quality of turker's work will be detected.  This protects against
malicious attackers who go ``undercover'' as one of our testers, only to turn
malicious and generate bad results when presented with real test profiles. 

\begin{figure*}[t]
\begin{minipage}{0.32\textwidth}
 \centering
	\includegraphics[width=1\textwidth]{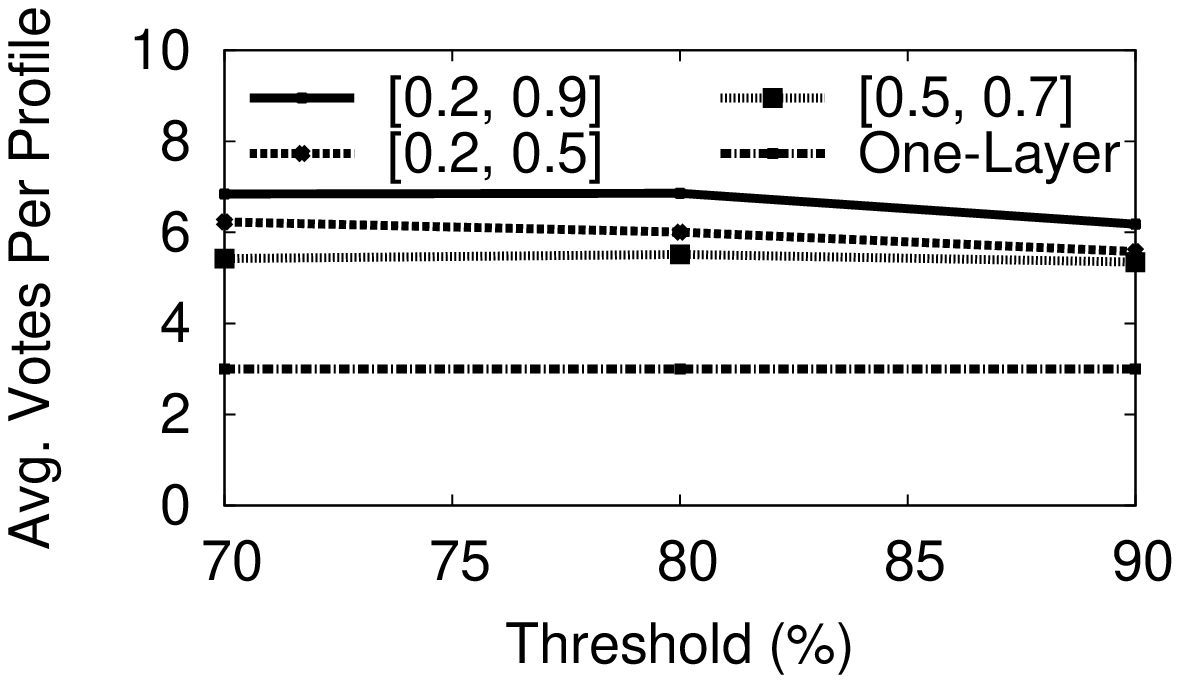}
	\caption{Threshold versus average votes per profiles. }
	\label{fig:thre-cost}
\end{minipage}
\hfill
\begin{minipage}{0.32\textwidth}
 \centering
     	\includegraphics[width=1\textwidth]{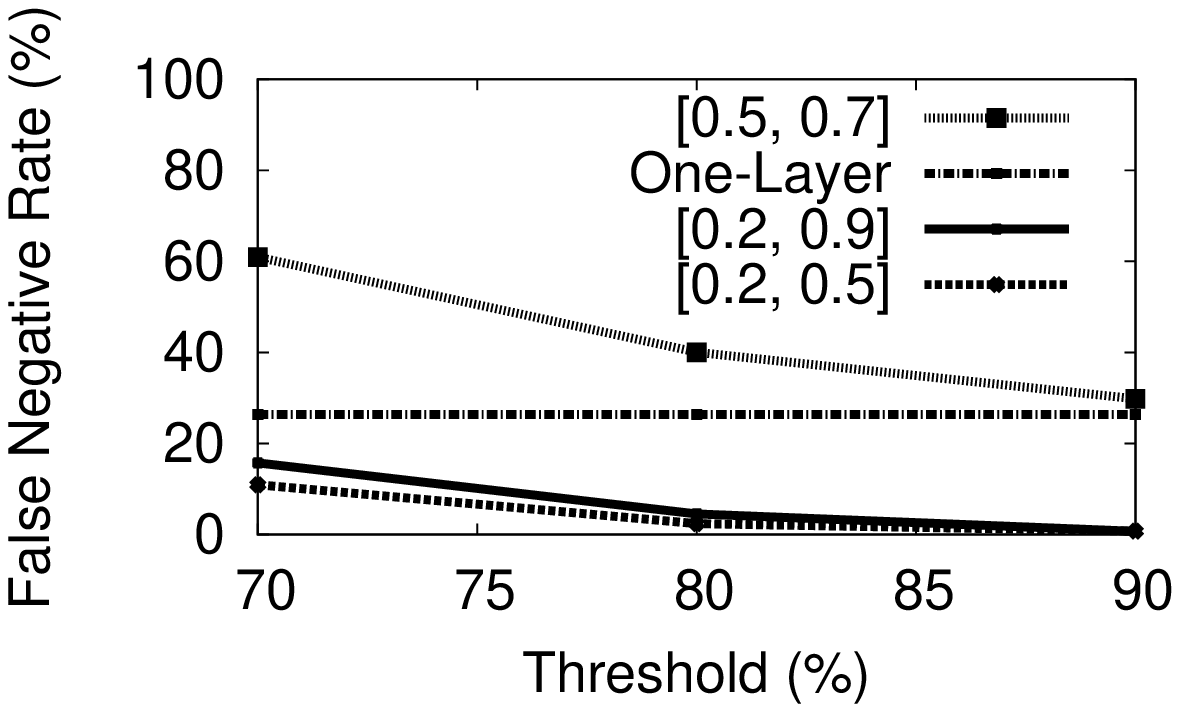}
	\caption{Threshold versus false negative rate. }
	\label{fig:thre-fn}
\end{minipage}
\hfill
\begin{minipage}{0.32\textwidth}
 \centering
	\includegraphics[width=1\textwidth]{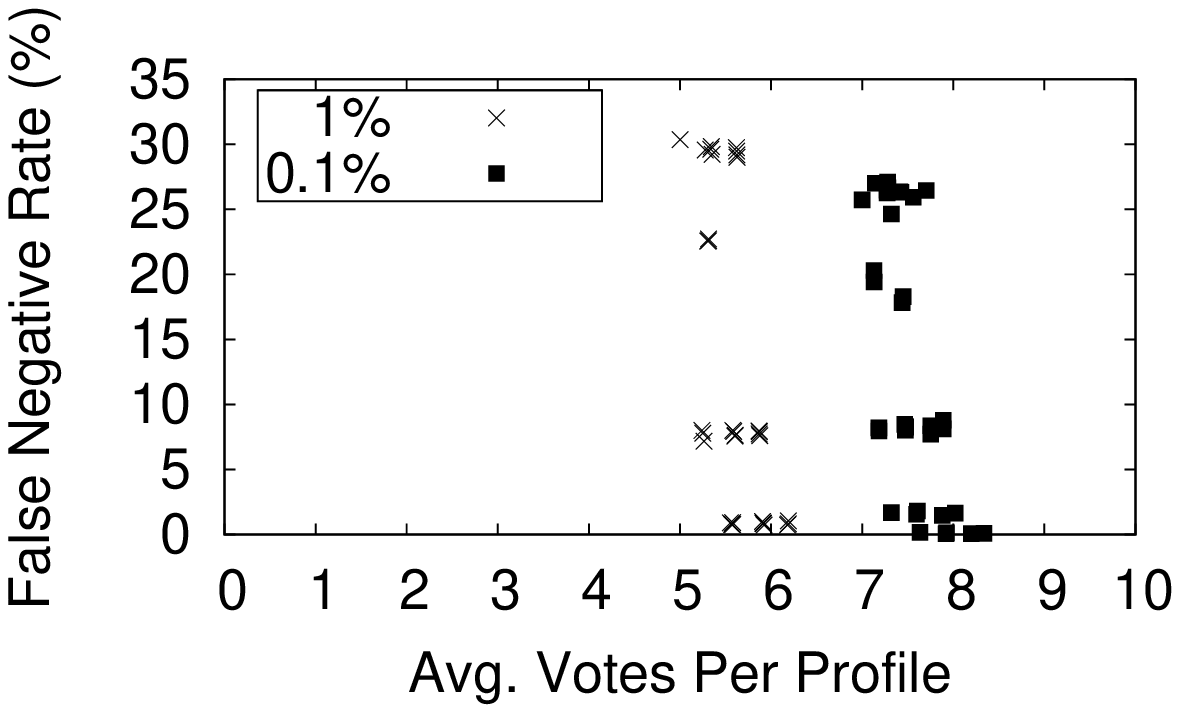}
	\caption{Tradeoff between votes per profile and desired accuracy. }
	\label{fig:cost-fn}
\end{minipage}
\end{figure*}

\subsection{System Simulations and Accuracy}
\compact
In this section, we examine the tradeoff between accuracy and cost in our
system. The overall goal of the system is to minimize false positives and
negatives, while also minimizing the number of votes needed per profile
(since each vote from a turker costs money). There is a clear tradeoff
between these two goals: as shown in Figure~\ref{fig:cn-wnum-fpfn}, more
votes reduces false positives.

\para{Simulation Methodology.} We use trace-driven simulations to examine
these tradeoffs.  We simulate 2000 suspicious profiles (1000 Sybil, 1000
legitimate) that need to be evaluated. We vary the number of votes per
profile, $V$, and calculate the false positive and false negative rates of
the system. Thus, each profile is evaluated by $V$ random turkers, and each
turker's probability of being correct is based on their results from our user
study. In keeping with our system design, all turkers with $<$60\% accuracy
are eliminated before the simulation by our ``turker selection'' module.

We consider two different ways to organize turkers in our simulations. The
first is simple: all turkers are grouped together. We refer
to this as {\em one-layer} organization. We also consider {\em two-layer}
organization: turkers are divided into two groups, based on an accuracy threshold
$T$. Turkers with accuracy $>T$ are placed in the {\em upper layer}, otherwise they
go into the {\em lower layer}.

In the two-layer scheme, profiles are first evaluated by turkers in the lower layer.
If there is strong consensus among the lower layer that the profile
is Sybil or legitimate, then the classification stands. However, if the
profile is {\em controversial}, then it is sent to the more accurate, upper
layer turkers for reevaluation. Each profile receives $B$ votes in
the lower layer and $U$ votes in the upper layer.
Intuitively, the two-layer system tries to maximize the utility of
the very accurate turkers by only having them evaluate difficult profiles.
Figure~\ref{fig:detector} depicts the two-layer version of our system.

In our design, we cap the maximum acceptable false positive rate at 1\%.  Our
motivation is obvious: social network providers will not deploy a security
system that has a non-negligible negative impact on legitimate users.  We
conducted simulations on all our turker groups, with consistent results
across groups.  For brevity, we limit our discussion here to results for the
Chinese turkers. As shown in Figure~\ref{fig:worker-fpfn}, the Chinese
turkers have the worst overall accuracy of our turker groups. Thus, they
represent the worst-case scenario for our system. The US and Indian groups
both exhibit better performance in terms of cost and accuracy during
simulations of our system.

\begin{table}
\begin{small}
\begin{center}
\scalebox{0.95}{
\begin{tabular}{|c||c|c|c|}
\hline
Threshold & 70\%  & 80\% & 90\%\\
\hline
$L$ (Lower Layer, Accurate Turkers) & 5 & 5 & 5 \\
\hline
$U$ (Upper Layer, Very Accurate Turkers)  & 3 & 3 & 2 \\ 
\hline
\end{tabular}}
\caption{Optimal \# of votes per profile in each layer in order to
  keep the false positives $<$1\%.}
\label{tab:sim-votes}
\end{center}
\end{small}
\vspace{-0.1in}
\end{table}

\para{Votes per Profile.}  In the one-layer simulations, the only variable is
votes per profile $V$. Given our constraint on false positives $<$1\%, we use
multiple simulations to compute the minimum value of $V$.  The simulations
reveal that the minimum number of votes for the Chinese profiles is 3; we use
this value in the remainder of our analysis.

Calculating the votes per profile in the two-layer case is more complicated,
but still tractable. The two-layer scenario includes four variables: votes per profile
($U$ upper and $L$ lower), the accuracy threshold between the layers $T$, and the
controversial range $R$ in which profiles are forwarded from the lower to upper layer.
To calculate $L$ and $U$ we use the same methodology as in Figure~\ref{fig:cn-wnum-fpfn}.
Turkers are divided into upper and lower layers for a given threshold $T \in [70\%,90\%]$,
then we incrementally increase the votes per profile in each layer until the false
positive rate is $<$1\%. The false positive rate of each layer is independent (\ie the
number of votes in the lower layer does not impact
votes in the upper layer), which simplifies the calculations. The controversial
range only effects the false negative rate, and is ignored from these calculations.

Table~\ref{tab:sim-votes} shows the minimum number of votes per profile needed in
the upper and lower layers as $T$ is varied. We use these values in the remainder
of our analysis.

Figure~\ref{fig:thre-cost} shows the average votes per profile in our simulations.
Three of the lines represent two-layer simulations with different $R$ values. For
example, $R=[0.2,0.9]$ means that if between 20\% and 90\% of the turkers classify
the profile as a Sybil, then the profile is considered controversial. Although we
simulated many $R$ ranges, only three representative ranges are shown for clarity.
The number of votes for the one-layer scheme is also shown.

The results in Figure~\ref{fig:thre-cost} show that the number of votes needed
in the various two-layer scenarios are relatively stable. As $R$ varies, the number
of profiles that must be evaluated by {\em both} layers changes. Thus,
average votes per profile fluctuates, although the average is always $\le L + U$
from Table~\ref{tab:sim-votes}. Overall, these fluctuations are minor, with average
votes only changing by $\approx$1.

\para{False Negatives.} Judging by the results in Figure~\ref{fig:thre-cost},
the one-layer scheme appears best because it requires the fewest votes per profile
(and is thus less costly). However, there is a significant tradeoff for lowering the cost
of the system: more false negatives.

Figure~\ref{fig:thre-fn} shows the false negative rates for our simulations.
The results for the two-layer scheme are superior: for certain values of $R$ and
thresholds $\ge$80\%, two-layers can achieve false negative rates $<$10\%. The
parameters that yield the lowest false negatives (0.7\%) and the
fewest average votes per profile (6) are $R = [0.2,0.5]$ and $T = 90\%$. We use
these parameters for the remainder of our analysis.

The results in Figures~\ref{fig:thre-cost} and~\ref{fig:thre-fn} capture
the power of our crowdsourced Sybil detection system. Using only an average
of 6 votes per profile, the system products results with false positive and negative rates 
both below 1\%. 

\para{Reducing False Positives.} In some situations, a social network may want to achieve
a false positive rate significantly lower than 1\%.  In order to evaluate how much this change
would affect costs, we re-ran all our simulations with the target false positive
rate set to $<$0.1\%. Figure~\ref{fig:cost-fn} plots the number of votes per
profile versus false negatives as the target false positive rate is varied. Each
point in the scatter is a different combination of $R$ and $T$ values. The
conclusion from Figure~\ref{fig:cost-fn} is straightforward: to get $<$0.1\% false
positives, you need two additional votes per turker. This tradeoff is fairly
reasonable: costs increase 33\%, but false positives reduce by an order of
magnitude.

\para{Parameterization.}  Since our system parameters were optimized using
actual user test results, they may not be ideal for every system or user
population. The key takeaway is that given a user
population, the system can be calibrated to provide high accuracy and
scalability.  We do not have sufficient data to make conjectures about how
often or when systems require re-calibration, but it is likely that a
deployed system might periodically recalibrate 
parameters such as $V$ and $T$ for continued accuracy.

\subsection{The Costs of a Turker Workforce}
\compact
Using the parameters derived in the previous section, we can
estimate how many turkers would be needed to deploy our system. Using
the parameters for Renren, each profile requires 6 votes on average, and
turkers can evaluate one profile every 20 seconds (see Figure~\ref{fig:acc-time}).
Thus, a turker working a standard 8-hour day (or several turkers working
an equivalent amount of time) can examine 1440 profiles.

Data from a real OSN indicates that the number of turkers needed for our
system is reasonable. According to~\cite{sybilrank},
Tuenti, a Spanish online social network, has a user-base of 11 million and
averages 12,000 user reports 
per day.  Our system would require 50 full-time turkers to handle this load.
If we scale the population size and reports per day up by a factor of 10, we
can estimate the load for a large OSN like Facebook. In this case, our system
requires 500 turkers. Our own experience showed that recruiting this
many turkers is not difficult (Table~\ref{tab:dataset}).  In fact, following our
crowdsourcing experiments on this and other projects~\cite{crowdturf-www12},
we received numerous messages from crowd requesting more tasks to perform.

Finally, we estimate the monetary cost of our system. Facebook pays turkers
from oDesk \$1 per hour to moderate images~\cite{oDeskface}. If we assume the
same cost per hour per turker for our system, then the daily cost for
deployment on Tuenti (\ie 12,000 reports per day) would only be \$400. This
compares favorably with Tuenti's existing practices: Tuenti pays 14 full-time
employees to moderate content~\cite{sybilrank}.  The estimated annual salary
for Tuenti employees are roughly
\euro30,000\footnote{\url{http://www.glassdoor.com/Salary/Tuenti-Salaries-E245751.htm}},
which is about \$20 per hour. So the Tuenti's moderation cost is \$2240 per
day, which is significantly more than the estimated costs of our turker
workforce.

\subsection{Privacy}
\compact
Protecting user privacy is a challenge for crowdsourced Sybil detection. How do you let
turkers evaluate user profiles without violating the privacy of those users? This issue
does not impact our experiments, since all profiles are from public accounts. However,
in a real deployment, the system needs to handle users with strict privacy settings.

One possible solution is to only show turkers the public portions of users' profiles.
However, this approach is problematic because Sybils could hinder the detection system
by setting their profiles to private.  Setting the profile to private may make it more
difficult for Sybils to friend other users, but it also cripples the discriminatory
abilities of turkers.

A better solution to the privacy problem is to leverage the OSNs existing
``report'' filter.  Suppose Alice reports Bob's profile as malicious. The
turker would be shown Bob's profile {\em as it appears to Alice}.
Intuitively, this gives the turker access to the same information that Alice
used to make her determination.  If Alice and Bob are friends, then the
turker would also be able to access friend-only information.  On the other
hand, if Alice and Bob are strangers, then the turker would only have access
to Bob's public information.  This scheme prevents users from abusing the
report system to leak the information of random strangers.

\section{Related Work}
\compact

The success of crowdsourcing platforms on the web has generated a great deal
of interest from researchers. Several studies have measured aspects of Amazon's
Mechanical Turk, including worker demographics~\cite{amazon-blog,demographic-chi10}
and task pricing~\cite{pricing-aaai11,amazonanalysis-acm,financial-hcomp09}.
There are studies that explore the pros and cons to use MTurk for user study~\cite{userstudies-chi08}.

Many studies address the problem of how to maximize accuracy from
inherently unreliable turkers. The most common approach is to use majority
voting~\cite{snow-emnlp08, Le-SIGIR10}, although this scheme is vulnerable to
collusion attacks by malicious turkers~\cite{Sun-HCI11}. Another approach is to
pre-screen turkers with a questionnaire to filter out less reliable
workers~\cite{gold-HCI11}. Finally, \cite{Sun-HCI11} proposes using a
tournament algorithm to determine the correct answer for difficult tasks.

In this study, we propose using crowdsourcing to solve a challenging OSN
security problem. However, many studies have demonstrated how crowdsourcing
can be used by attackers for malicious ends. Studies have observed
malicious HITs asking turkers to send social spam~\cite{crowdturf-www12},
perform search engine optimization (SEO)~\cite{freelance}, write fake
reviews~\cite{ott-EtAl}, and even install malware on their
systems~\cite{malware-woot}.

\section{Conclusion and Open Questions}

Sybil accounts challenge the stability and security of today's online social
networks.  Despite significant efforts from researchers and industry,
malicious users are creating increasingly realistic Sybil accounts that blend
into the legitimate user population.  To address the problem today, social
networks rely on ad hoc solutions such as manual inspection by employees.

Our user study takes the first step towards the development of a scalable and
accurate crowdsourced Sybil detection system.  Our results show that by
using experts to calibrate ground truth filters, we can eliminate low
accuracy turkers, and also separate the most accurate turkers from the crowd.
Simulations show that a hierarchical two-tiered system can both be accurate and
scalable in terms of total costs.

\para{Ground-truth.} Our system evaluation is constrained by the
ground-truth Sybils used in our user study, \ie it is possible that there
are additional Sybils that were not caught and included in our data. Thus,
our results are a lower bound on detection accuracy. Sybils that can bypass
Facebook or Renren's existing detection mechanisms could potentially be
caught by our system.

\para{Deployment.} Effective deployment of crowdsourced Sybil detection
mechanisms remains an open question.  We envision that the crowdsourcing
system will be used to complement existing techniques such as
content-filtering and statistical models.  For example, output from accurate
turkers can teach automated tools which fields of the data can most easily
identify fake accounts.  Social networks can further lower the costs of
running this system by utilizing their own users as crowdworkers. The social
network can replace monetary payments with in-system virtual currency, {\em
  e.g.} Facebook Credits, Zynga Cash, or Renren Beans.  We are currently
discussing internal testing and deployment possibilities with collaborators
at Renren and LinkedIn.

\para{Countermeasures.}  An effective solution must take into account
possible countermeasures by attackers. For example, ground-truth profiles
must be randomly mixed with test profiles in order to detect malicious
turkers that attempt to poison the system by submitting intentionally inaccurate
results. The ground-truth profiles must be refreshed periodically to avoid
detection.  In addition, it is possible for attackers to infiltrate the system in
order to learn how to improve fake profiles to avoid detection. Dealing with
these ``undercover'' attackers remains an open question.



\newpage
\balance
\bibliographystyle{latex8}
\bibliography{zhao,astro}

\end{document}